\documentclass[aps,pra,preprint,showpacs,amsmath,amssymb,floatfix,nofootinbib,superscriptaddress,10pt]{revtex4-1} 
\usepackage[latin1]{inputenc}  
\usepackage{graphicx}        
\usepackage{bm}              
\usepackage{subfigure}
\usepackage{xcolor}
\usepackage{hyperref}

\begin{document}

\title{Genetic optimization of attosecond pulse generation in light-field synthesizers}

\author{E. Balogh}
\thanks{These authors contributed equally to the results.}
\affiliation{Department of Optics and Quantum Electronics, University of Szeged, H-6720 Szeged, Hungary}
\author{B. B\'odi}
\thanks{These authors contributed equally to the results.}
\affiliation{MTA Lend\"ulet Ultrafast Nanooptics Group, Wigner Research Centre for Physics, H-1121 Budapest, Hungary}
\author{V. Tosa}
\affiliation{National Institute for R\&D of Isotopic and Molecular Technologies, RO-400293 Cluj-Napoca, Romania}
\author{E. Goulielmakis}
\affiliation{Max-Planck-Institut f\"ur Quantenoptik, 85748 Garching, Germany}
\author{K. Varj\'u}
\affiliation{Department of Optics and Quantum Electronics, University of Szeged, H-6720 Szeged, Hungary}
\affiliation{ELI-HU Nonprofit Kft., 6720 Szeged, Hungary}
\author{P. Dombi}
\affiliation{MTA Lend\"ulet Ultrafast Nanooptics Group, Wigner Research Centre for Physics, H-1121 Budapest, Hungary}
\affiliation{Max-Planck-Institut f\"ur Quantenoptik, 85748 Garching, Germany}
\affiliation{ELI-HU Nonprofit Kft., 6720 Szeged, Hungary}

\date{\today}

\begin{abstract}

We demonstrate control over attosecond pulse generation and shaping by numerically optimizing the synthesis of few-cycle to sub-cycle driver waveforms. 
The optical waveform synthesis takes place in an ultrabroad spectral band covering the ultraviolet-infrared domain. 
These optimized driver waves are used for ultrashort single and double attosecond pulse production (with tunable separation) revealing the potentials of the light wave synthesizer device demonstrated by Wirth et al. [Science \textbf{334}, 195 (2011)].
The results are also analyzed with respect to attosecond pulse propagation phenomena.

\end{abstract}

\maketitle

\section{Introduction}

The rapid development of femtosecond laser technology provided the tool for studying extreme nonlinear optical interaction of laser light with atoms resulting in  high-order harmonic generation (HHG).
Thus the production of attosecond pulse trains \cite{farkas,Paul01062001} and isolated attosecond light pulses \cite{2001SAP,2001attometro,krauszrevmod} became possible.
These, being the shortest controllably producible, coherent light pulses available today, are used in basic research to study the time-resolved evolution of electron wave packets during phenomena like photo-ionization \cite{Schultze25062010,2010twophoton,2011photoion,2011PRLStreak}, hole migration \cite{2006Remacle}, Auger decay \cite{Drescher2002,2012Auger} and others occurring at the attosecond time scale \cite{2006electroninterferometry,2010realtime}.
Since the generation of attosecond light pulses is a highly nonlinear process, it is very sensitive to the properties of the generating (driver) pulse.
This poses strict requirements on the shot-to-shot stability of the energy, shape and carrier-envelope phase (CEP) of the generating pulse. 
On the other hand, this sensitivity also offers an opportunity to produce attosecond pulses with widely varying parameters, by fine-tuning the properties of the generating laser pulse.

The recently developed light-field synthesizers \cite{Wirth2011,2012LFS} made possible the production of few-cycle laser pulses with almost arbitrary shape.
This promises unique possibilities for highly-controlled attosecond pulse production.
However, from the practical point-of-view, one needs to consider the large number of degrees of freedom in a light-field synthesizer. 
Such a device is based on the spectral separation of laser radiation to 3--4 interferometer channels where amplitudes, CEP values and relative delays are independently adjustable. 
After the recombination of these channels to a single-cycle or sub-cycle optical driver pulse, the high-harmonic generation process can take place. 
Attosecond pulses are typically generated by spectral and spatial filtering of this high-harmonic beam \cite{krauszrevmod}.
The two or three independently controllable parameters for each spectral channel, together with spectral filtering of the harmonic beam, generate a parameter space with so many dimensions, that it is already infeasible scan it experimentally in order to study the capabilities of the light-field synthesizer setup in attosecond pulse production.

In this paper we analyze the possibilities offered by light-field synthesizers in attosecond pulse generation using a genetic algorithm.
These types of algorithms, widely used for optimization of many-parameter processes involving highly nonlinear response functions, enable the search for a parameter set that would produce a predefined result, in our case, attosecond pulse(s) with specific features.

As attosecond pulses are used in pump-probe experiments to study electron dynamics at the shortest possible time scales, the main goals of research in this area are to produce even shorter light pulses \cite{80as,Zhao2012}, to realize XUV pump XUV probe experiments \cite{2011xuvpp}, and to extend their applicability by increasing the photon energy \cite{2010NPhotPopmintchev,2012kev}, photon flux \cite{2011Sansone,2013nctakahashi} and repetition rate \cite{2012khz,2012incavity,2013mhzhhg} of the generated harmonics.

Evolutionary algorithms have been used in numerical calculations of HHG to optimize harmonic yield \cite{2001roos,2001hhgoptim,2012OCT} and phase matching \cite{2001PRLChristov}, extend the cutoff \cite{2009cutoff,2013cutoffext}, and to generate single attosecond pulses SAPs \cite{2004sapoptimization,Tang2010155}.
The optimization of several experimental HHG setups has also been carried out with self-learning algorithms \cite{2000Bartels,2004Bartels,2008RevModPhys,raey}.
In order to study the possibilities offered by light-field synthesizers in attosecond pulse generation, in our numerical optimization we set the goals to produce the shortest possible isolated attosecond pulses, and attosecond double-pulses with variable separation between them.

\section{Simulation methods}

\begin{figure}[htb!]%
\centering
	\includegraphics[width=0.45\textwidth]{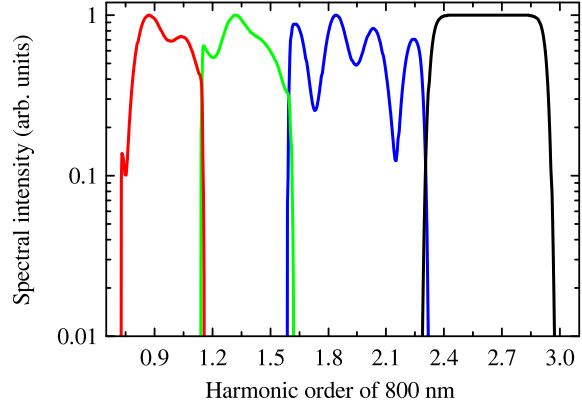}
\caption{(Color online) Spectral channels of the driving wave from 1040 nm to 272 nm. The boundary wavelengths correspond to 698 nm, 501 nm and 347 nm. We extended the three experimentally demonstrated channels with a fourth super-Gaussian in the UV range.}%
\label{fig1}%
\end{figure}

Our goal was to implement a simulation environment based on an experimentally demonstrated synthesizer instrument \cite{Wirth2011}, in order to optimize its parameters numerically and to check the possibilities of different atto-pulse shape generations by tailoring the driving pulse. 
The modeled device divides a supercontinuum into three channels (Fig.\ref{fig1}) covering 1.51 octaves.
A delay line, wedge pairs and an aperture are built into each spectral channel making it possible to control of the three pulses' delay, CEP and energy, respectively. 
Finally these are recombined to form an extremely short (single-cycle or sub-cycle) laser pulse \cite{Wirth2011}.

\begin{figure*}[htb!]%
	\includegraphics[width=\textwidth]{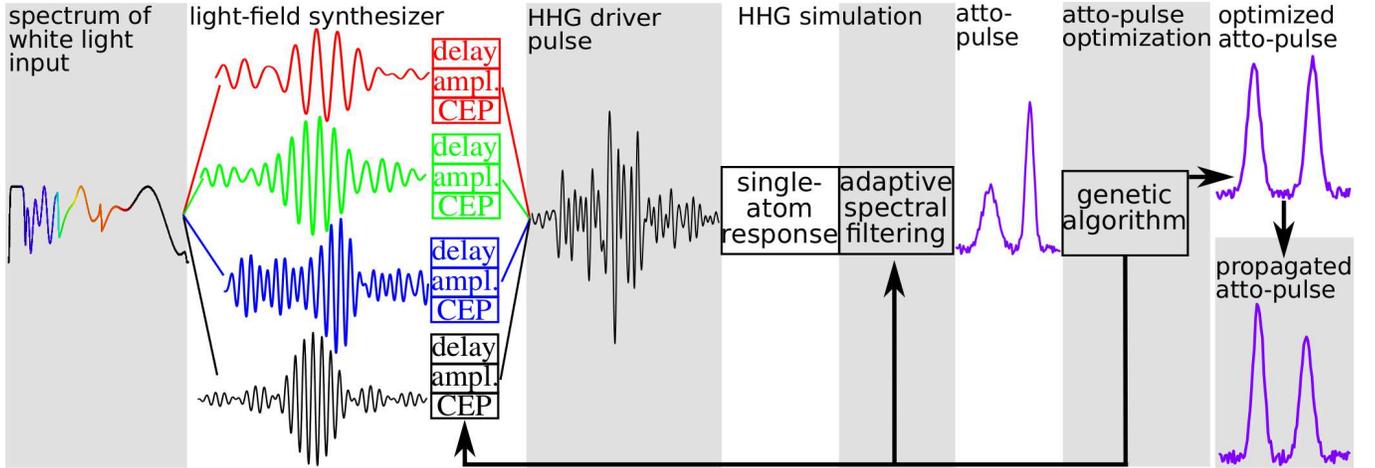}
\caption{(Color online) Flowsheet of the genetic optimization process of a light-field synthesizer for the generation of tailored attosecond pulses.}%
\label{fig2}%
\end{figure*}

Aiming to have more freedom, in the simulation, in addition to the three channels with experimental spectra, we added a fourth (UV) channel extending the total usable bandwidth to 1.88 octaves, and tuned the parameters of each channel to have more control over the synthesized driving field.
The choice of the spectral region of the fourth super-Gaussian channel was based on recent developments of the field synthesizer \cite{lefteris}.
In our model a fine delay, the CEP, and the amplitude can be set for all four channels (Fig.\ref{fig2}), defining a set of independent parameters.
After adding the channels together again, the driver was normalized to a constant maximum. 
The simulation then calculates the Lewenstein integral for the synthesized driver waveform \cite{lewenstein} as a single-atom response in Ne gas.

The optimization part was done by a genetic algorithm; this is an adequate choice because of the high nonlinearity of the result in the parameters. 
In each case a fitness function is defined to decide how suitable the individual driver pulse is for predefined simulation goals (such as short attosecond pulse generation or double attosecond pulse generation). 
Starting with a random population the code eliminates the worst individuals in every generation, and by dynamically adjusting the mutation ranges, it converges in 50--80 generations.

The response of a single atom to the synthesized driver field is calculated in the strong-field approximation \cite{lewenstein,lewph} which, for the time-dependent dipole moment yields:
\begin{align}
x(t) &= i\int_{-\infty}^t dt' \Big( \frac{\pi}{\epsilon+i(t-t')/2} \Big)^{3/2}d^{\ast}[p_{st}(t',t)-A(t)]  \nonumber\\
     & d[p_{st}(t',t)-A(t')] E(t') *exp[-iS_{st}(t',t)] + c.c. \nonumber\\
\label{eq:lew}
\end{align}
where $E(t)$ and $A(t)$ denote the time dependent electric field and vector potential of the laser pulse, $d(p)$ denotes the atomic dipole matrix element for the bound-free transition, $\epsilon$ being a small positive number to remove the divergence at $t=t'$, while $p_{st}$ is the stationary point of the canonical momentum, and $S(t',t)$ is the quasi-classical action.

The harmonic spectral amplitudes are calculated as $\omega^2 x(\omega)$, where $x(\omega)$ is the Fourier transform of the time-dependent dipole moment.

Since the generated harmonic radiation has an intrinsic chirp (that is so robust that even with the genetic optimization it could not be changed substantially), the attosecond pulse obtained from the full high-harmonic spectrum is not the shortest pulse. 
To find the conditions of shortest pulse generation we used spectral filtering in the genetic algorithm as an extra degree of freedom before calculating the inverse Fourier transform (iFT).
The spectral limits used in the iFT are also parameters of the genetic optimization, hence they are defined by the algorithm and may change in each optimization step.

It is known that each harmonic component of the spectrum has contributions from two significant trajectories per half optical cycle, called the short and long trajectories.
The radiation produced by the two trajectories has different properties, and as a result, the long trajectory components are usually eliminated in experiments either naturally by phase matching or by spatial filtering of the harmonic beam.
In most cases this elimination is also desirable, as this way the harmonic emission becomes less divergent, the attosecond pulses become shorter and moreover, the developed post-compression methods make use of the initial positive chirp, characteristic of short trajectory radiation only \cite{lopezprl}.

In semi-classical models of HHG, these two trajectories differ in their travel time from ionization to recombination.
When calculating equation \ref{eq:lew} the lower $t'$ and upper $t$ limits of the integral define the longest electron travel time that is accounted for in the model. By limiting the integration to a short time region, the emissions from recombining electrons that perform long trajectories can be minimized.
Inserting a full macroscopic model into a genetic optimization process (calculating tens of generations before convergence) exceeds the computing capacity available for most research groups, so we chose to emulate the macroscopic elimination of long trajectories in the genetic algorithm by restricting the integration time in equation \ref{eq:lew} so that only short trajectory components contribute to the dipole spectrum.
Since the instantaneous frequency of the synthesized laser field practically changes in each half cycle, we performed classical calculations to find the travel time corresponding to the cutoff trajectory, and the obtained value was set as integration time.

In order to establish to what extent the limitation of the integration time can be used to mimic part of the macroscopic effects for the complicated fields in question, we performed three-dimensional (3D) calculations using the driver fields optimized by the genetic algorithm for single-atom interaction, and compared the results to the integral-limited single-atom calculations.
In the 3D model we assume Gaussian beams with 40 $\mu$m beam waist for all four fields, and a gas medium placed at their common focus.
In these calculations the effect of dispersion, absorption, plasma dispersion and the optical Kerr effect are taken into account when propagating the four pulses through the gas cell. 
The wave equations for the four fields are solved as described in \cite{tosamodel} where the nonlinear terms contributing to the refractive indices (optical Kerr effect and plasma dispersion) are calculated from the superposition of the four fields.
In this way the four fields propagate independently in the same medium but each field "sees" a specific refractive index as given by the nonlinear terms noted previously.
For the ionization the Ammosov--Delone--Krainov model is used \cite{adk}, and the ionization by harmonic field through re-absorption is neglected.

The propagated fields are then used to calculate of the single-atom response over the interaction region, now without restricting the integration time in equation \ref{eq:lew}.
We assume linear polarizations in a common direction for all fields, thus the single-atom response is also linearly polarized in the same direction and enters the source term for the wave equation describing the propagation of the harmonic field. 
The final result of this step is the harmonic near field at the exit of the gas medium.
The far field is eventually obtained using Huygens' integral for arbitrary \textit{ABCD} ray-transfer matrices \cite{siegman}, assuming that the harmonic beam is propagated through an aperture and refocused, a procedure which eliminates the components with high divergence.

We note here that the ionization model used by us looses accuracy when applied to few-cycle pulses \cite{2000adk}, therefore we tried to minimize the role of ionization in our interpretation of the macroscopic results.
To this end, no optimization of the macroscopic parameters has been carried out, apart from selecting a gas pressure which favors the phase-matching of short trajectories (in our case this varies from 20 to 66 mbar), something that is usually the case in experiments as well.
The length of the gas cell is also limited to just 0.6 mm in all cases, in order to minimize the role of macroscopic effects.

\section{Results}

\subsection{Isolated attosecond pulses}

Today the shortest SAP demonstrated experimentally is 67 as long, produced with double optical gating and compressed by a zirconium filter. 
That setup uses polarization gating to isolate a SAP from a pulse train, and advantageous macroscopic effects to eliminate high group delay dispersion harmonics close to the cutoff that would make the generated pulse longer \cite{Zhao2012}.
Without the use of the polarization gating technique, the shortest SAP demonstrated is 80 as long, measured after passing through a zirconium foil used for post-compression \cite{80as}.
The length of the generated pulse is defined by its bandwidth and phase locking of the constituting spectral components -- both of which are improved by increasing the driver intensity \cite{2003Mairesse}.  
Since the aim of this paper is not to find the achievable pulse duration minimum but to study the extent of tunability of the high-harmonic generation process by exotic driver field shapes, in the optimization algorithm we fixed the peak intensity of the generated pulse to $1.5\times10^{15}$ W/cm$^2$. 
At this intensity the generating sub-cycle pulse produces an ionization rate of $\sim$15--22\% (depending on the pulse shape) in neon, which is still feasible to be used experimentally.

If we take a driver pulse that can be derived from the light-field synthesizer spectrum (Fig. 1.) with constant spectral phase (i.e. no optimization performed) and normalized to the same peak intensity, we can generate a SAP with a 73 as FWHM pulse length with only the single-atom response considered. 
Therefore, we set out to see first whether a genetic optimization process of the driving field can deliver a better result.

\begin{figure*}[htb!]%
	\subfigure{
	\includegraphics[width=0.4\textwidth]{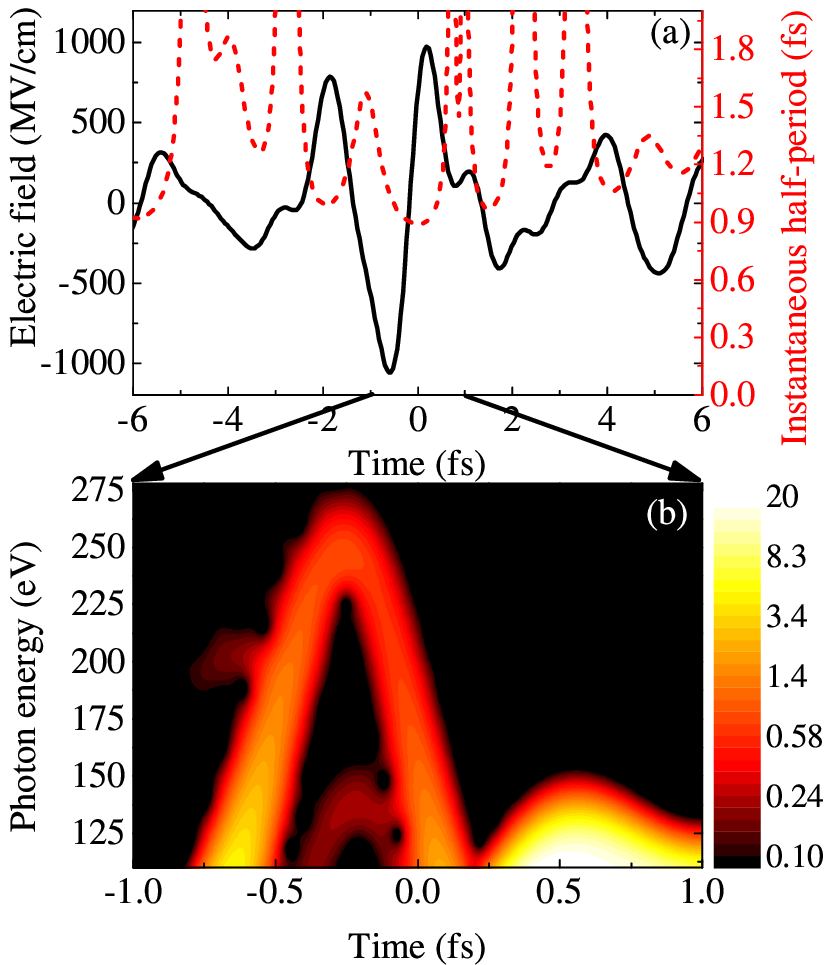}}
	\subfigure{
	\includegraphics[width=0.4\textwidth]{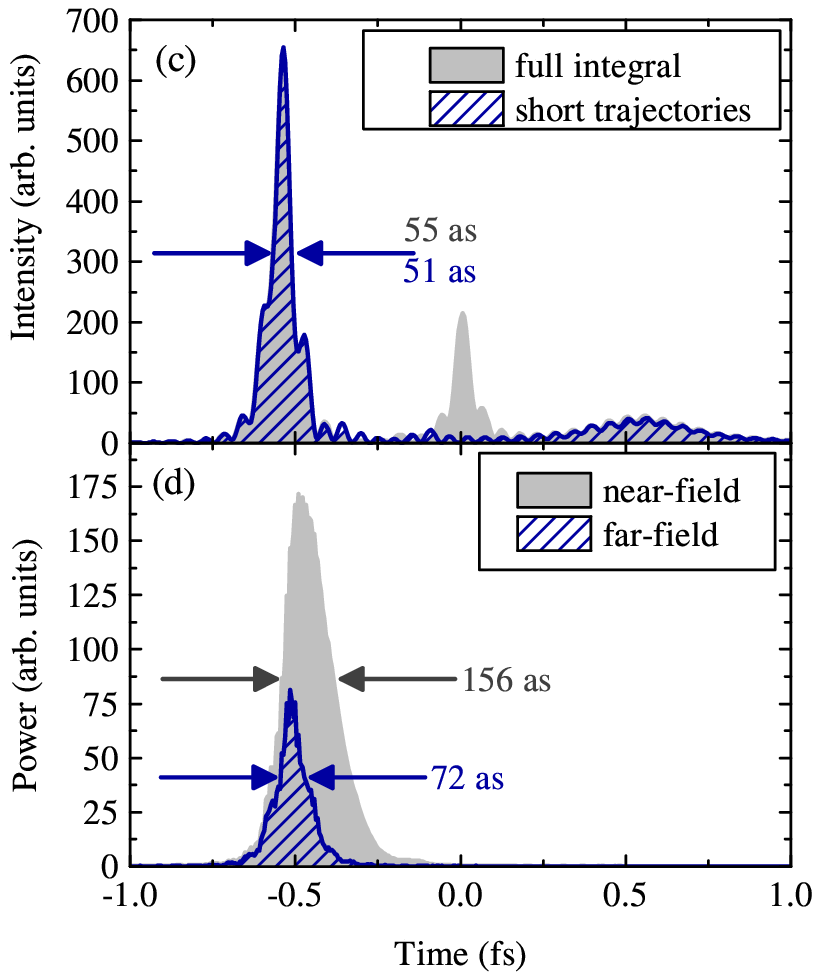}}
\caption{(Color online) Generation of short, isolated attosecond pulses. (a) Driver electric field (black, solid line) and instantaneous half-cycle period (red, dashed line). (b) Time-frequency map of the generated dipole radiation from the full Lewenstein integral shows the presence of the two trajectory components, (c) which results in the two pulses in the single-atom calculations. (d) Macroscopic effects, however, eliminate the second pulse, and spatial filtering also makes the remaining one shorter. Harmonics between 118 and 195 eV are used here for the synthesis of the attosecond pulse.}%
\label{fig3}%
\end{figure*}

Our first goal was to apply the genetic algorithm to find the optimum field shape producing the shortest possible single attosecond pulse with at least 10:1 contrast ratio, varying all the available free parameters, with the restraint of the fixed peak intensity. 
The genetic algorithm converged to a solution, where the optimized single-atom calculations predicts the possibility to generate 55 as long SAP (see Fig.\ref{fig3}).
The optimized laser field is illustrated by its electric field and the half-cycle period ($ \pi / |d\varphi / dt| $) (part a). 
Gabor transform of the single-atom Lewenstein dipole (part b) indicates the time-frequency structure of the emitted radiation. 
We observe a broad emission between -1.0 and 0.2 fs, and there is also a narrower emission around 0.5 fs emitted in the next half-cycle, however this is eliminated from the temporal picture by spectral filtering.
The optimization procedure indicated 118 eV to 195 eV bandwidth for the shortest single pulse production and an average GDD of 1460 as$^2$.
The spectrally filtered radiation transformed to the time domain (part c) indicates 51 as pulses when the Lewenstein integral is limited to include only short trajectories. 
The unlimited integral produces a second peak corresponding to long trajectories cf. time-frequency map.
Macroscopic calculations predicts slightly longer 72 as pulse duration with the same generating field, when radiation with $>$1 mrad divergence is filtered out of the harmonic beam (Fig.\ref{fig3}(d)). 
We add here, that the 73 as pulses generated by the constant spectral phase driving field (cosine shaped) would stretch to a 106 as length by propagation, and is narrowed to 95 as by spatial filtering.

The other criterion used in this optimization, having a SAP with a minimum contrast ratio of 1:10 (that is, suppressing the side pulses), is also fulfilled and improved in the macroscopic results. 
We see that the contribution of long trajectories and the second side-pulse is already eliminated by the end of the gas cell, but the spatial filtering is required to shorten the pulse in the far-field. 
We find that the improved phase-locking achieved by filtering out harmonics generated off-axis \cite{2002SpaceTime} is responsible for the shortening of the pulse from the 156 as in the near-field to the 72 as measurable in the far-field.
The pulse durations quoted in this section are as-generated values, i.e. not using any postcompression method. 
Alternatively, a different optimization procedure can be carried out when postcompression is also available and the inherent chirp is not limiting the achievable pulse duration.

\subsection{Double attosecond pulses with variable separation}

\begin{figure*}[htbp!]%
	\subfigure{
	\includegraphics[width=0.4\textwidth]{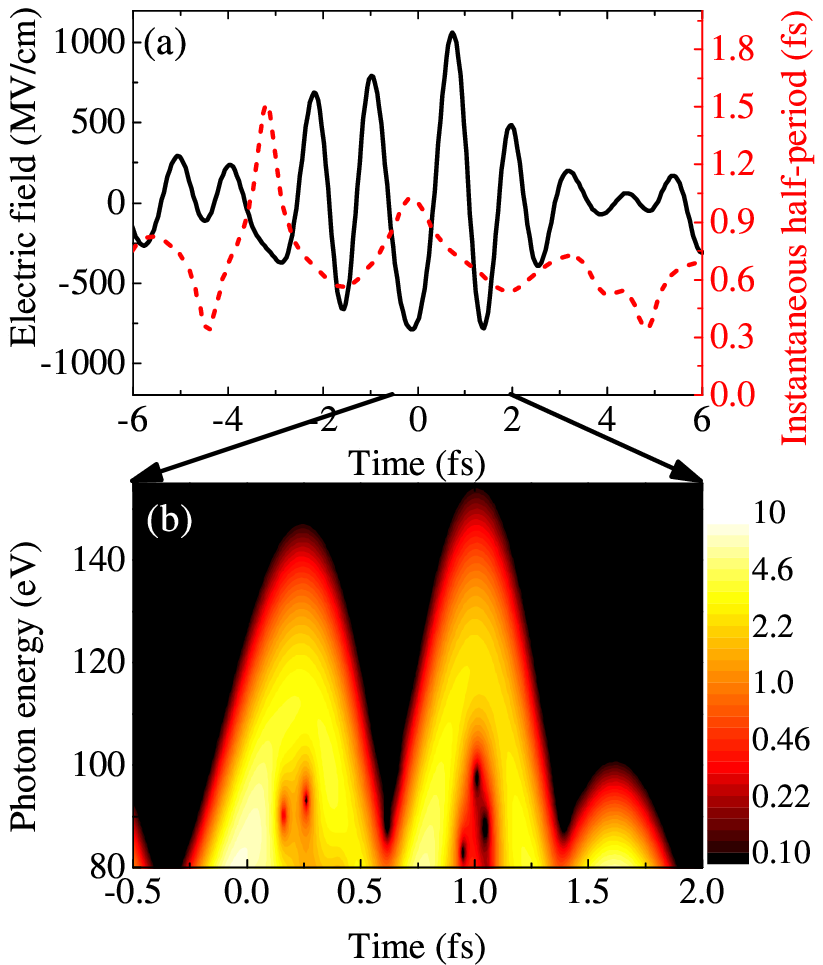}}
	\subfigure{
	\includegraphics[width=0.4\textwidth]{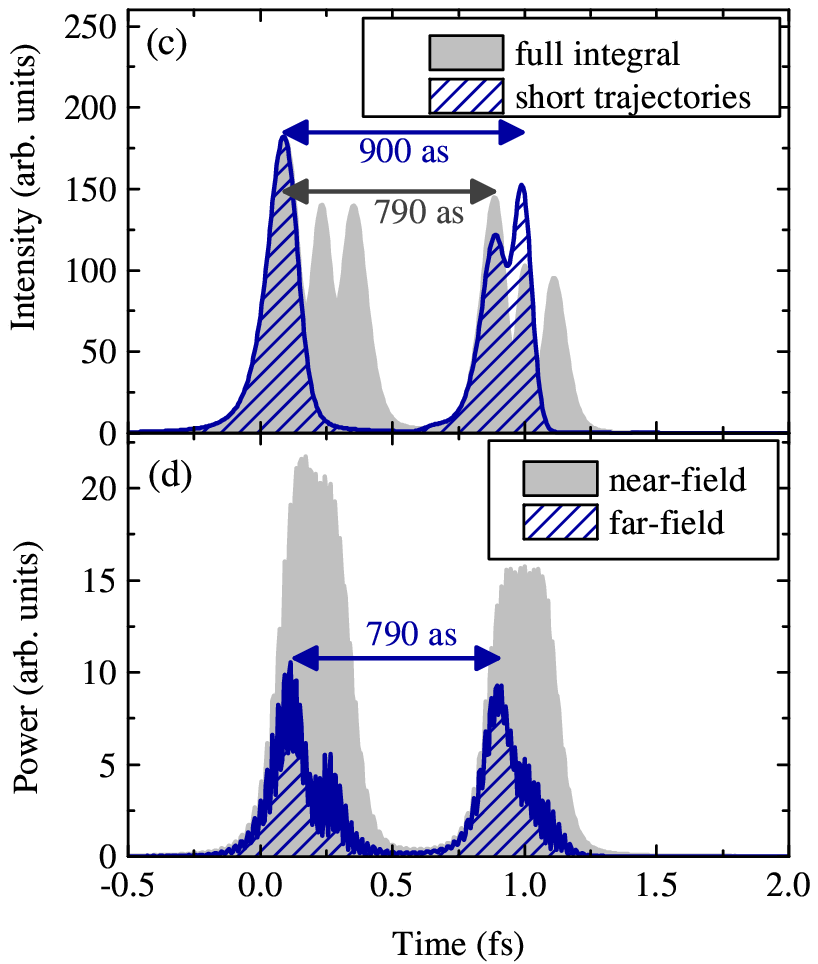}}
\caption{(Color online) Double pulse, with 900 as separation. (a) The electric field (black, continuous line) and instantaneous half-period (red, dashed line) calculated from the phase derivative shows the complex structure of the driving field. (b) Time-frequency analysis showing the two pulses created in different half-cycles. Suppressing harmonics below 81 eV produces the two pulses which are present both in (c) single-atom and (d) macroscopic results.}%
\label{fig4}%
\end{figure*}

\begin{figure*}[htbp!]%
	\subfigure{
	\includegraphics[width=0.4\textwidth]{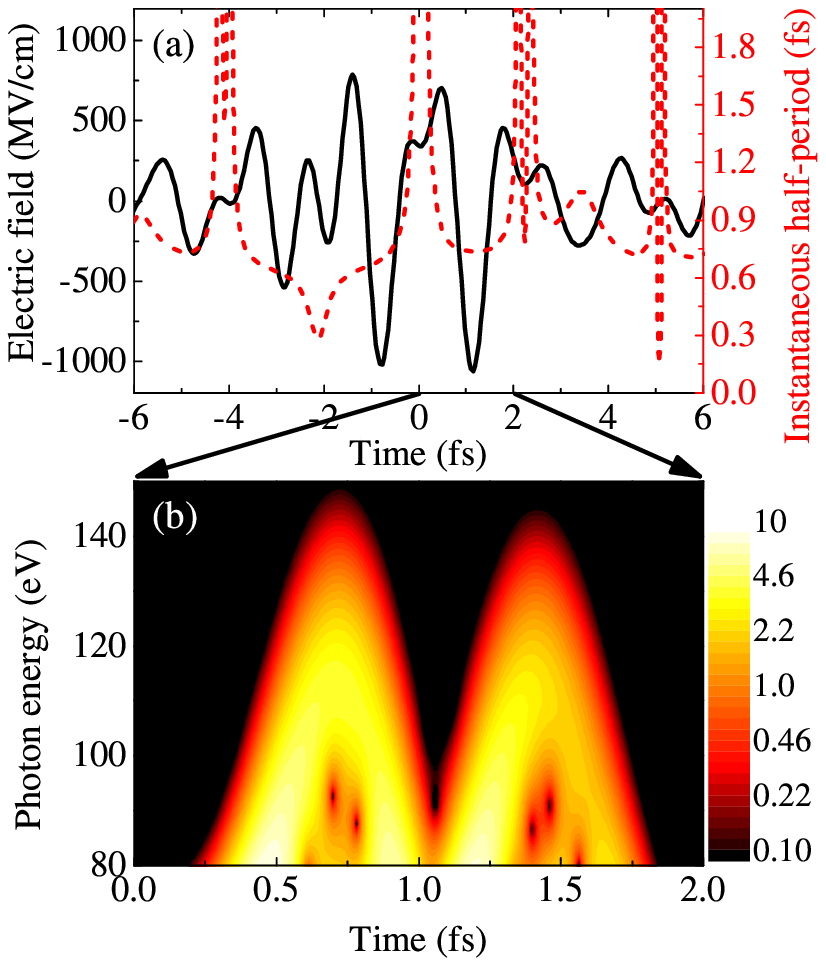}}
	\subfigure{
	\includegraphics[width=0.4\textwidth]{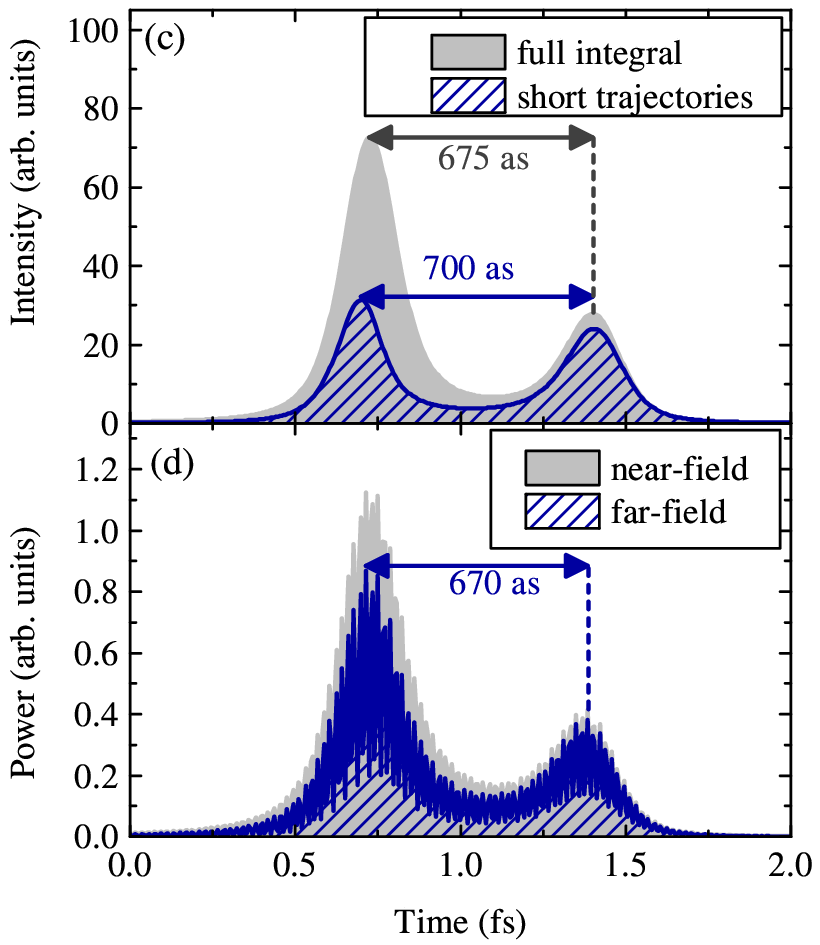}}
\caption{(Color online) Double pulse, with 700 as separation. (a) Generating electric field (black, solid line) and instantaneous half-period (red, dashed line). (b) Time-frequency analysis showing the two pulses created in different half-cycles. Spectral selection from 108 to 170 eV produces the two pulses which are present in both (c) single-atom and (d) macroscopic results.}%
\label{fig5}%
\end{figure*}

The generation of double attosecond pulses (DAP) with variable separation on the multi-femtosecond scale has been predicted to be realizable using multi-cycle driver pulses and the polarization gating technique \cite{Miao2012}. 
Normally the natural separation of pulses in an attosecond pulse train is rooted in the half-cycle periodicity of the process: 1.3 fs pulse separation for the 800 nm fundamental. 
In the present work we focused on producing DAPs with sub-femtosecond separation: aiming at equally intense pulses with 900, 700, 500 and 300 as separation. 
The results of the optimization process are illustrated in Figs \ref{fig4}--\ref{fig5} in the same manner as Fig. \ref{fig3}. 


For 900 and 700 as separations (Fig.\ref{fig4} and Fig.\ref{fig5}) single-atom calculations yielded the required pulse structure, and the results were also confirmed by macroscopic calculations, although the separation of the pulses is slightly smaller and their amplitudes are slightly different in the macroscopic response for both cases. 
Our analysis shows that for both cases the two pulses are produced by short trajectory components in different half-cycles of the generating driver field. 

We could not link the central frequency of the driver pulse to the separation present between the two pulses, as this frequency corresponds to half-periods of 700 and 705 as in both cases mentioned above.
We also see, that the instantaneous half period of the generating fields varies between 900 as and 600 as in the interval where the production (ionization -- free travel -- recombination of the electrons) of the attosecond pulses takes place. 
Due to these sub-cycle variations in the driver pulse, we cannot directly link the instantaneous period either to the separation of the pulses, emphasizing that the desired results are produced by non-trivial field shapes.

\begin{figure*}[htb!]%
	\subfigure{
	\includegraphics[width=0.4\textwidth]{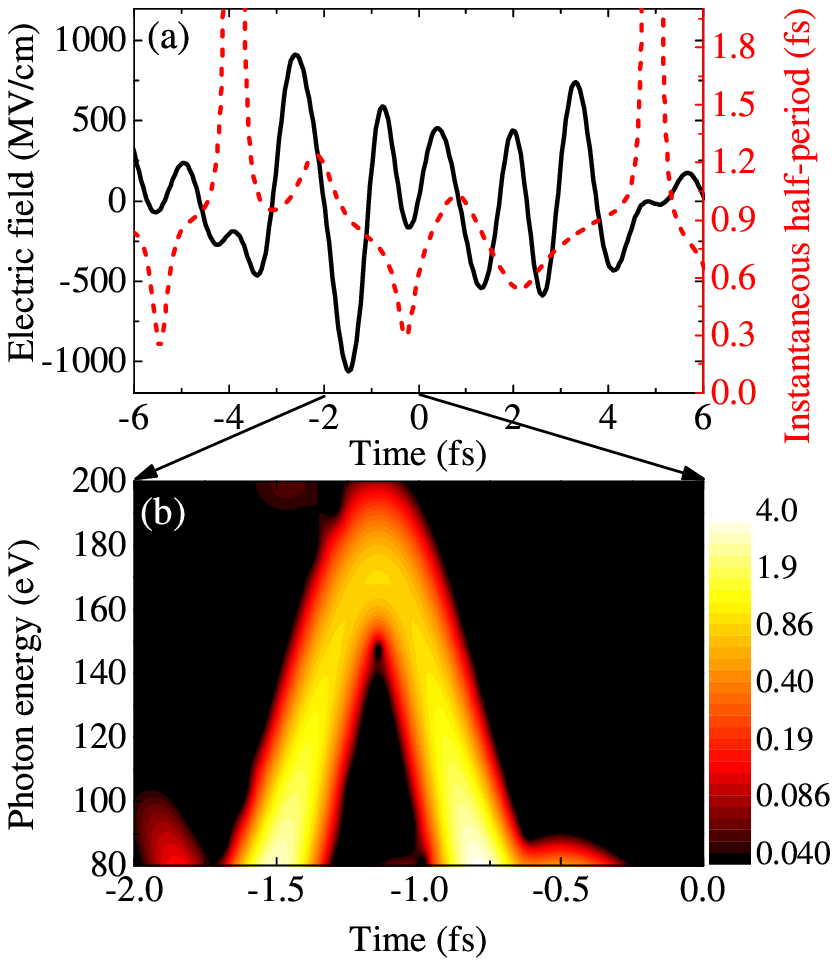}}
	\subfigure{
	\includegraphics[width=0.4\textwidth]{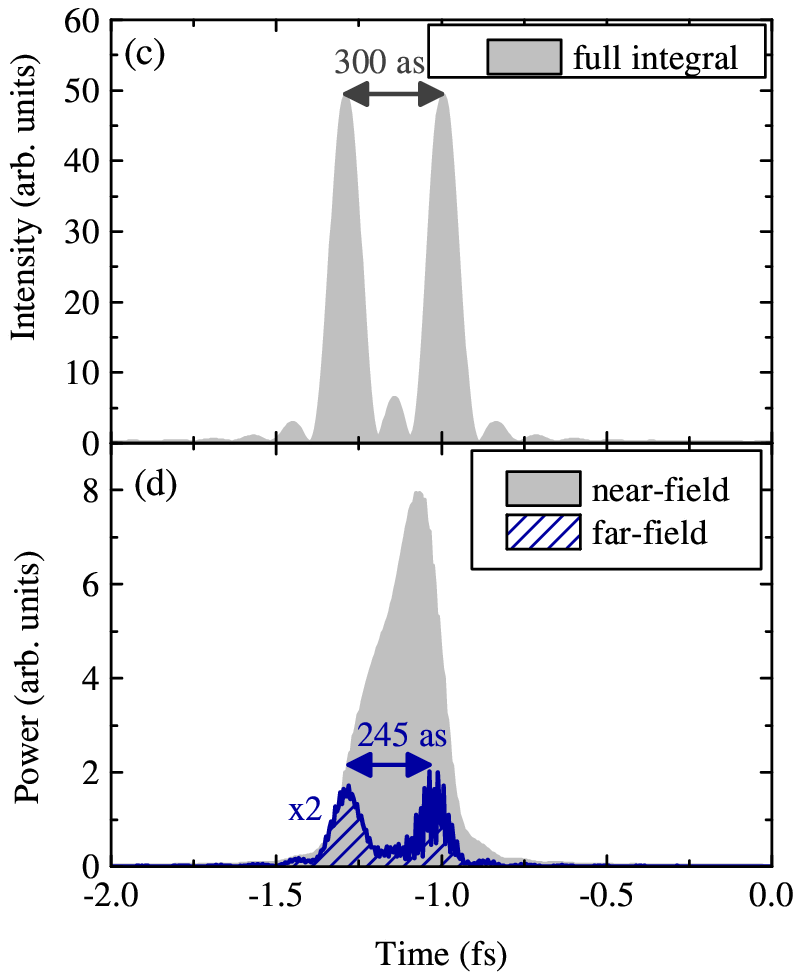}}
\caption{(Color online) Double pulse, with 300 as separation. (a) Generating electric field (black, solid line) and instantaneous half-period (red, dashed line). (b) Time-frequency analysis showing the two pulses created in the same half-cycle by short and long trajectories. (c) Spectral selection (from 130 to 166 eV) produces the two pulses in the single-atom calculations. (d) These merge in the near-field, but are again separated by spatial filtering of the harmonic beam.}%
\label{fig6}%
\end{figure*}

On the other hand, we found that the generation of attosecond double pulses with less separation can be performed in a fundamentally different manner. 
When optimizing the driver wave for attosecond pulse production with 300 as distance, we ended up with two trajectory sets emitting harmonics in the same half-cycle (Fig.\ref{fig6}).
The separation between attosecond pulses generated from short and long trajectories can be fine-tuned with the applied spectral filters \cite{chang}: the separation becomes smaller when the spectral window is moved towards higher harmonics, i.e. towards harmonics closer to the cutoff, where the short and long trajectories merge.

In the case shown in Fig.\ref{fig6} the two pulses merge in the near-field, and spatial filtering is essential to separate them. 
This also means that a large part (90\%) of the pulse energy is filtered out. 
In this configuration, the fact that a lower pressure has been used -- favoring phase-matching of long trajectory components -- also helps to keep both pulses in the far field with comparable peak power.
At this point we also mention that in the results presented here the effect of gas pressure was not significant when it was kept between the limits mentioned earlier (20 and 66 mbar).
The only exception is this last case (DAP with 300 as separation), where the role of long trajectories is significant, and single-atom results could only be reproduced by using the lower gas pressure of 20 mbar.

It is seen that short (300 as) and long (700 and 900 as) separation of the attosecond pulses were realizable via different schemes. 
With the spectral components for the driver laser field shown in Fig.\ref{fig1}, our algorithm produces double attosecond pulses with separation larger than 660 as from two different half-cycles of the driver laser pulse.
We find that shorter separations are possible only from short and long trajectory components of harmonic radiation generated in the same half-cycle.
Since we observe no direct relationship between separation and field shape, the use of the genetic algorithm is necessary to obtain the required pulse separation.

With the optimization tools at hand, we also examined the possibility of the generation of non-trivial attosecond pulse shapes with the help of the light-field synthesizer setup, such as top hat or triangular attosecond pulses. 
In these cases, however, no convergence was observed and these exotic pulse shapes seem to be out of reach in the attosecond pulse generation process.

\section{Conclusions}
We applied a genetic algorithm to optimize gas high-harmonic generation in a modeled multivariable light-field synthesizer device. 
We inserted only the single-atom response -- with limited temporal integration -- in the optimization cycle due to processor time limitations, but afterwards each selected driver waveform was checked with a 3D macroscopic model. 
We found that whereas an unoptimized light-field synthesizer delivers 73 as pulses, the genetic optimization delivers driver waveforms that produce 55 as pulses remaining robust after propagation as well. 
Therefore, such a non-trivial optimization of the light-field synthesizer parameters is essential for the production of the shortest possible pulses allowed by this sophisticated experimental setup.
The generation process was also optimized to produce double attosecond pulses with tunable sub-femtosecond separation. 
For pulse separations above 660 as the genetic algorithm returned field shapes where consecutive half-cycles produce the two pulses, whereas for a displacement as short as 300 as, we obtained double pulse generation via short and long trajectory components within a single half-cycle.

We note that in the present work the parameters of the macroscopic model have not been optimized. 
Not to accentuate propagation effects, a regular scenario has been selected with a short cell, positioned right after the focus, low gas pressure (20--66 mbar) and a common beam waist of 40 $\mu$m for all four fields.
Genetic algorithms have proved to be useful in optimizing macroscopic conditions (not including optimization of the generating field) \cite{2001roos}, therefore we expect even better results if both optimizations are performed simultaneously (e.g with supercomputing facilities).
With automatic characterization of the generated pulses, and motorized tuning of the light-field synthesizer, the genetic algorithm might be used to optimize the generation process in vivo.

\section{Acknowledgments}
We wish to acknowledge fruitful discussions with G. Farkas. 
The project was supported by the European Community's FP7 Programme under contract ITN-2008-238362 (ATTOFEL) and by the Hungarian Scientific Research Fund (OTKA project 81364 and NN107235).
The project was partially funded by ``TAMOP-4.2.2.A-11/1/KONV-2012-0060 - Impulse lasers for use in materials science and biophotonics'' supported by the European Union and co-financed by the European Social Fund.
VT acknowledges partial support from contract PN-II-PCE-2012-4-0342.
EG acknowledges support by European Research Council grant (Attoelectronics-258501) and the Deutsche Forschungsgemeinschaft Cluster of Excellence: Munich Centre for Advanced Photonics (\url{http://www.munich-photonics.de/}).
KV and PD acknowledge support from the Bolyai and ``Lend\"ulet'' Grants of the Hungarian Academy of Sciences and the Partner Group Program of the Max Planck Society.

\bibliography{literatureg}

\begin{thebibliography}{50}%
\makeatletter
\providecommand \@ifxundefined [1]{%
 \@ifx{#1\undefined}
}%
\providecommand \@ifnum [1]{%
 \ifnum #1\expandafter \@firstoftwo
 \else \expandafter \@secondoftwo
 \fi
}%
\providecommand \@ifx [1]{%
 \ifx #1\expandafter \@firstoftwo
 \else \expandafter \@secondoftwo
 \fi
}%
\providecommand \natexlab [1]{#1}%
\providecommand \enquote  [1]{``#1''}%
\providecommand \bibnamefont  [1]{#1}%
\providecommand \bibfnamefont [1]{#1}%
\providecommand \citenamefont [1]{#1}%
\providecommand \href@noop [0]{\@secondoftwo}%
\providecommand \href [0]{\begingroup \@sanitize@url \@href}%
\providecommand \@href[1]{\@@startlink{#1}\@@href}%
\providecommand \@@href[1]{\endgroup#1\@@endlink}%
\providecommand \@sanitize@url [0]{\catcode `\\12\catcode `\$12\catcode
  `\&12\catcode `\#12\catcode `\^12\catcode `\_12\catcode `\%12\relax}%
\providecommand \@@startlink[1]{}%
\providecommand \@@endlink[0]{}%
\providecommand \url  [0]{\begingroup\@sanitize@url \@url }%
\providecommand \@url [1]{\endgroup\@href {#1}{\urlprefix }}%
\providecommand \urlprefix  [0]{URL }%
\providecommand \Eprint [0]{\href }%
\providecommand \doibase [0]{http://dx.doi.org/}%
\providecommand \selectlanguage [0]{\@gobble}%
\providecommand \bibinfo  [0]{\@secondoftwo}%
\providecommand \bibfield  [0]{\@secondoftwo}%
\providecommand \translation [1]{[#1]}%
\providecommand \BibitemOpen [0]{}%
\providecommand \bibitemStop [0]{}%
\providecommand \bibitemNoStop [0]{.\EOS\space}%
\providecommand \EOS [0]{\spacefactor3000\relax}%
\providecommand \BibitemShut  [1]{\csname bibitem#1\endcsname}%
\let\auto@bib@innerbib\@empty
\bibitem [{\citenamefont {Farkas}\ and\ \citenamefont {Toth}(1992)}]{farkas}%
  \BibitemOpen
  \bibfield  {author} {\bibinfo {author} {\bibfnamefont {G.}~\bibnamefont
  {Farkas}}\ and\ \bibinfo {author} {\bibfnamefont {C.}~\bibnamefont {Toth}},\
  }\href {\doibase 10.1016/0375-9601(92)90534-S} {\bibfield  {journal}
  {\bibinfo  {journal} {Physics Letters A}\ }\textbf {\bibinfo {volume}
  {168}},\ \bibinfo {pages} {447} (\bibinfo {year} {1992})}\BibitemShut
  {NoStop}%
\bibitem [{\citenamefont {Paul}\ \emph {et~al.}(2001)\citenamefont {Paul},
  \citenamefont {Toma}, \citenamefont {Breger}, \citenamefont {Mullot},
  \citenamefont {Aug\'{e}}, \citenamefont {Balcou}, \citenamefont {Muller},\
  and\ \citenamefont {Agostini}}]{Paul01062001}%
  \BibitemOpen
  \bibfield  {author} {\bibinfo {author} {\bibfnamefont {P.~M.}\ \bibnamefont
  {Paul}}, \bibinfo {author} {\bibfnamefont {E.~S.}\ \bibnamefont {Toma}},
  \bibinfo {author} {\bibfnamefont {P.}~\bibnamefont {Breger}}, \bibinfo
  {author} {\bibfnamefont {G.}~\bibnamefont {Mullot}}, \bibinfo {author}
  {\bibfnamefont {F.}~\bibnamefont {Aug\'{e}}}, \bibinfo {author}
  {\bibfnamefont {P.}~\bibnamefont {Balcou}}, \bibinfo {author} {\bibfnamefont
  {H.~G.}\ \bibnamefont {Muller}}, \ and\ \bibinfo {author} {\bibfnamefont
  {P.}~\bibnamefont {Agostini}},\ }\href {\doibase 10.1126/science.1059413}
  {\bibfield  {journal} {\bibinfo  {journal} {Science}\ }\textbf {\bibinfo
  {volume} {292}},\ \bibinfo {pages} {1689} (\bibinfo {year}
  {2001})}\BibitemShut {NoStop}%
\bibitem [{\citenamefont {Drescher}\ \emph {et~al.}(2001)\citenamefont
  {Drescher}, \citenamefont {Hentschel}, \citenamefont {Kienberger},
  \citenamefont {Tempea}, \citenamefont {Spielmann}, \citenamefont {Reider},
  \citenamefont {Corkum},\ and\ \citenamefont {Krausz}}]{2001SAP}%
  \BibitemOpen
  \bibfield  {author} {\bibinfo {author} {\bibfnamefont {M.}~\bibnamefont
  {Drescher}}, \bibinfo {author} {\bibfnamefont {M.}~\bibnamefont {Hentschel}},
  \bibinfo {author} {\bibfnamefont {R.}~\bibnamefont {Kienberger}}, \bibinfo
  {author} {\bibfnamefont {G.}~\bibnamefont {Tempea}}, \bibinfo {author}
  {\bibfnamefont {C.}~\bibnamefont {Spielmann}}, \bibinfo {author}
  {\bibfnamefont {G.~A.}\ \bibnamefont {Reider}}, \bibinfo {author}
  {\bibfnamefont {P.~B.}\ \bibnamefont {Corkum}}, \ and\ \bibinfo {author}
  {\bibfnamefont {F.}~\bibnamefont {Krausz}},\ }\href {\doibase
  10.1126/science.1058561} {\bibfield  {journal} {\bibinfo  {journal}
  {Science}\ }\textbf {\bibinfo {volume} {291}},\ \bibinfo {pages} {1923}
  (\bibinfo {year} {2001})}\BibitemShut {NoStop}%
\bibitem [{\citenamefont {Hentschel}\ \emph {et~al.}(2001)\citenamefont
  {Hentschel}, \citenamefont {Kienberger}, \citenamefont {Spielmann},
  \citenamefont {Reider}, \citenamefont {Milosevic}, \citenamefont {Brabec},
  \citenamefont {Corkum}, \citenamefont {Heinzmann}, \citenamefont {Drescher},\
  and\ \citenamefont {Krausz}}]{2001attometro}%
  \BibitemOpen
  \bibfield  {author} {\bibinfo {author} {\bibfnamefont {M.}~\bibnamefont
  {Hentschel}}, \bibinfo {author} {\bibfnamefont {R.}~\bibnamefont
  {Kienberger}}, \bibinfo {author} {\bibfnamefont {C.}~\bibnamefont
  {Spielmann}}, \bibinfo {author} {\bibfnamefont {G.}~\bibnamefont {Reider}},
  \bibinfo {author} {\bibfnamefont {N.}~\bibnamefont {Milosevic}}, \bibinfo
  {author} {\bibfnamefont {T.}~\bibnamefont {Brabec}}, \bibinfo {author}
  {\bibfnamefont {P.}~\bibnamefont {Corkum}}, \bibinfo {author} {\bibfnamefont
  {U.}~\bibnamefont {Heinzmann}}, \bibinfo {author} {\bibfnamefont
  {M.}~\bibnamefont {Drescher}}, \ and\ \bibinfo {author} {\bibfnamefont
  {F.}~\bibnamefont {Krausz}},\ }\href {\doibase 10.1038/35107000} {\bibfield
  {journal} {\bibinfo  {journal} {Nature}\ }\textbf {\bibinfo {volume} {414}},\
  \bibinfo {pages} {509} (\bibinfo {year} {2001})}\BibitemShut {NoStop}%
\bibitem [{\citenamefont {Krausz}\ and\ \citenamefont
  {Ivanov}(2009)}]{krauszrevmod}%
  \BibitemOpen
  \bibfield  {author} {\bibinfo {author} {\bibfnamefont {F.}~\bibnamefont
  {Krausz}}\ and\ \bibinfo {author} {\bibfnamefont {M.}~\bibnamefont
  {Ivanov}},\ }\href {\doibase 10.1103/RevModPhys.81.163} {\bibfield  {journal}
  {\bibinfo  {journal} {Rev. Mod. Phys.}\ }\textbf {\bibinfo {volume} {81}},\
  \bibinfo {pages} {163} (\bibinfo {year} {2009})}\BibitemShut {NoStop}%
\bibitem [{\citenamefont {Schultze}\ \emph {et~al.}(2010)\citenamefont
  {Schultze}, \citenamefont {Fieß}, \citenamefont {Karpowicz}, \citenamefont
  {Gagnon}, \citenamefont {Korbman}, \citenamefont {Hofstetter}, \citenamefont
  {Neppl}, \citenamefont {Cavalieri}, \citenamefont {Komninos}, \citenamefont
  {Mercouris}, \citenamefont {Nicolaides}, \citenamefont {Pazourek},
  \citenamefont {Nagele}, \citenamefont {Feist}, \citenamefont {Burgd\"orfer},
  \citenamefont {Azzeer}, \citenamefont {Ernstorfer}, \citenamefont
  {Kienberger}, \citenamefont {Kleineberg}, \citenamefont {Goulielmakis},
  \citenamefont {Krausz},\ and\ \citenamefont {Yakovlev}}]{Schultze25062010}%
  \BibitemOpen
  \bibfield  {author} {\bibinfo {author} {\bibfnamefont {M.}~\bibnamefont
  {Schultze}}, \bibinfo {author} {\bibfnamefont {M.}~\bibnamefont {Fieß}},
  \bibinfo {author} {\bibfnamefont {N.}~\bibnamefont {Karpowicz}}, \bibinfo
  {author} {\bibfnamefont {J.}~\bibnamefont {Gagnon}}, \bibinfo {author}
  {\bibfnamefont {M.}~\bibnamefont {Korbman}}, \bibinfo {author} {\bibfnamefont
  {M.}~\bibnamefont {Hofstetter}}, \bibinfo {author} {\bibfnamefont
  {S.}~\bibnamefont {Neppl}}, \bibinfo {author} {\bibfnamefont {A.~L.}\
  \bibnamefont {Cavalieri}}, \bibinfo {author} {\bibfnamefont {Y.}~\bibnamefont
  {Komninos}}, \bibinfo {author} {\bibfnamefont {T.}~\bibnamefont {Mercouris}},
  \bibinfo {author} {\bibfnamefont {C.~A.}\ \bibnamefont {Nicolaides}},
  \bibinfo {author} {\bibfnamefont {R.}~\bibnamefont {Pazourek}}, \bibinfo
  {author} {\bibfnamefont {S.}~\bibnamefont {Nagele}}, \bibinfo {author}
  {\bibfnamefont {J.}~\bibnamefont {Feist}}, \bibinfo {author} {\bibfnamefont
  {J.}~\bibnamefont {Burgd\"orfer}}, \bibinfo {author} {\bibfnamefont {A.~M.}\
  \bibnamefont {Azzeer}}, \bibinfo {author} {\bibfnamefont {R.}~\bibnamefont
  {Ernstorfer}}, \bibinfo {author} {\bibfnamefont {R.}~\bibnamefont
  {Kienberger}}, \bibinfo {author} {\bibfnamefont {U.}~\bibnamefont
  {Kleineberg}}, \bibinfo {author} {\bibfnamefont {E.}~\bibnamefont
  {Goulielmakis}}, \bibinfo {author} {\bibfnamefont {F.}~\bibnamefont
  {Krausz}}, \ and\ \bibinfo {author} {\bibfnamefont {V.~S.}\ \bibnamefont
  {Yakovlev}},\ }\href {\doibase 10.1126/science.1189401} {\bibfield  {journal}
  {\bibinfo  {journal} {Science}\ }\textbf {\bibinfo {volume} {328}},\ \bibinfo
  {pages} {1658} (\bibinfo {year} {2010})}\BibitemShut {NoStop}%
\bibitem [{\citenamefont {Swoboda}\ \emph {et~al.}(2010)\citenamefont
  {Swoboda}, \citenamefont {Fordell}, \citenamefont {Kl\"under}, \citenamefont
  {Dahlstr\"om}, \citenamefont {Miranda}, \citenamefont {Buth}, \citenamefont
  {Schafer}, \citenamefont {Mauritsson}, \citenamefont {L'Huillier},\ and\
  \citenamefont {Gisselbrecht}}]{2010twophoton}%
  \BibitemOpen
  \bibfield  {author} {\bibinfo {author} {\bibfnamefont {M.}~\bibnamefont
  {Swoboda}}, \bibinfo {author} {\bibfnamefont {T.}~\bibnamefont {Fordell}},
  \bibinfo {author} {\bibfnamefont {K.}~\bibnamefont {Kl\"under}}, \bibinfo
  {author} {\bibfnamefont {J.~M.}\ \bibnamefont {Dahlstr\"om}}, \bibinfo
  {author} {\bibfnamefont {M.}~\bibnamefont {Miranda}}, \bibinfo {author}
  {\bibfnamefont {C.}~\bibnamefont {Buth}}, \bibinfo {author} {\bibfnamefont
  {K.~J.}\ \bibnamefont {Schafer}}, \bibinfo {author} {\bibfnamefont
  {J.}~\bibnamefont {Mauritsson}}, \bibinfo {author} {\bibfnamefont
  {A.}~\bibnamefont {L'Huillier}}, \ and\ \bibinfo {author} {\bibfnamefont
  {M.}~\bibnamefont {Gisselbrecht}},\ }\href {\doibase
  10.1103/PhysRevLett.104.103003} {\bibfield  {journal} {\bibinfo  {journal}
  {Phys. Rev. Lett.}\ }\textbf {\bibinfo {volume} {104}},\ \bibinfo {pages}
  {103003} (\bibinfo {year} {2010})}\BibitemShut {NoStop}%
\bibitem [{\citenamefont {Kl\"under}\ \emph {et~al.}(2011)\citenamefont
  {Kl\"under}, \citenamefont {Dahlstr\"om}, \citenamefont {Gisselbrecht},
  \citenamefont {Fordell}, \citenamefont {Swoboda}, \citenamefont {Gu\'enot},
  \citenamefont {Johnsson}, \citenamefont {Caillat}, \citenamefont
  {Mauritsson}, \citenamefont {Maquet}, \citenamefont {Taieb},\ and\
  \citenamefont {L'Huillier}}]{2011photoion}%
  \BibitemOpen
  \bibfield  {author} {\bibinfo {author} {\bibfnamefont {K.}~\bibnamefont
  {Kl\"under}}, \bibinfo {author} {\bibfnamefont {J.~M.}\ \bibnamefont
  {Dahlstr\"om}}, \bibinfo {author} {\bibfnamefont {M.}~\bibnamefont
  {Gisselbrecht}}, \bibinfo {author} {\bibfnamefont {T.}~\bibnamefont
  {Fordell}}, \bibinfo {author} {\bibfnamefont {M.}~\bibnamefont {Swoboda}},
  \bibinfo {author} {\bibfnamefont {D.}~\bibnamefont {Gu\'enot}}, \bibinfo
  {author} {\bibfnamefont {P.}~\bibnamefont {Johnsson}}, \bibinfo {author}
  {\bibfnamefont {J.}~\bibnamefont {Caillat}}, \bibinfo {author} {\bibfnamefont
  {J.}~\bibnamefont {Mauritsson}}, \bibinfo {author} {\bibfnamefont
  {A.}~\bibnamefont {Maquet}}, \bibinfo {author} {\bibfnamefont
  {R.}~\bibnamefont {Taieb}}, \ and\ \bibinfo {author} {\bibfnamefont
  {A.}~\bibnamefont {L'Huillier}},\ }\href {\doibase
  10.1103/PhysRevLett.106.143002} {\bibfield  {journal} {\bibinfo  {journal}
  {Phys. Rev. Lett.}\ }\textbf {\bibinfo {volume} {106}},\ \bibinfo {pages}
  {143002} (\bibinfo {year} {2011})}\BibitemShut {NoStop}%
\bibitem [{\citenamefont {Ivanov}\ and\ \citenamefont
  {Smirnova}(2011)}]{2011PRLStreak}%
  \BibitemOpen
  \bibfield  {author} {\bibinfo {author} {\bibfnamefont {M.}~\bibnamefont
  {Ivanov}}\ and\ \bibinfo {author} {\bibfnamefont {O.}~\bibnamefont
  {Smirnova}},\ }\href {\doibase 10.1103/PhysRevLett.107.213605} {\bibfield
  {journal} {\bibinfo  {journal} {Phys. Rev. Lett.}\ }\textbf {\bibinfo
  {volume} {107}},\ \bibinfo {pages} {213605} (\bibinfo {year}
  {2011})}\BibitemShut {NoStop}%
\bibitem [{\citenamefont {Remacle}\ and\ \citenamefont
  {Levine}(2006)}]{2006Remacle}%
  \BibitemOpen
  \bibfield  {author} {\bibinfo {author} {\bibfnamefont {F.}~\bibnamefont
  {Remacle}}\ and\ \bibinfo {author} {\bibfnamefont {R.~D.}\ \bibnamefont
  {Levine}},\ }\href {\doibase 10.1073/pnas.0601855103} {\bibfield  {journal}
  {\bibinfo  {journal} {PNAS}\ }\textbf {\bibinfo {volume} {103}},\ \bibinfo
  {pages} {6793} (\bibinfo {year} {2006})}\BibitemShut {NoStop}%
\bibitem [{\citenamefont {Drescher}\ \emph {et~al.}(2002)\citenamefont
  {Drescher}, \citenamefont {Hentschel}, \citenamefont {Kienberger},
  \citenamefont {Uiberacker}, \citenamefont {Yakovlev}, \citenamefont
  {Scrinzi}, \citenamefont {Westerwalbesloh}, \citenamefont {Kleineberg},
  \citenamefont {Heinzmann},\ and\ \citenamefont {Krausz}}]{Drescher2002}%
  \BibitemOpen
  \bibfield  {author} {\bibinfo {author} {\bibfnamefont {M.}~\bibnamefont
  {Drescher}}, \bibinfo {author} {\bibfnamefont {M.}~\bibnamefont {Hentschel}},
  \bibinfo {author} {\bibfnamefont {R.}~\bibnamefont {Kienberger}}, \bibinfo
  {author} {\bibfnamefont {M.}~\bibnamefont {Uiberacker}}, \bibinfo {author}
  {\bibfnamefont {V.}~\bibnamefont {Yakovlev}}, \bibinfo {author}
  {\bibfnamefont {A.}~\bibnamefont {Scrinzi}}, \bibinfo {author} {\bibfnamefont
  {T.}~\bibnamefont {Westerwalbesloh}}, \bibinfo {author} {\bibfnamefont
  {U.}~\bibnamefont {Kleineberg}}, \bibinfo {author} {\bibfnamefont
  {U.}~\bibnamefont {Heinzmann}}, \ and\ \bibinfo {author} {\bibfnamefont
  {F.}~\bibnamefont {Krausz}},\ }\href {\doibase 10.1038/nature01143}
  {\bibfield  {journal} {\bibinfo  {journal} {Nature}\ }\textbf {\bibinfo
  {volume} {419}},\ \bibinfo {pages} {803} (\bibinfo {year}
  {2002})}\BibitemShut {NoStop}%
\bibitem [{\citenamefont {Sch\"utte}\ \emph {et~al.}(2012)\citenamefont
  {Sch\"utte}, \citenamefont {Bauch}, \citenamefont {Fr\"uhling}, \citenamefont
  {Wieland}, \citenamefont {Gensch}, \citenamefont {Pl\"onjes}, \citenamefont
  {Gaumnitz}, \citenamefont {Azima}, \citenamefont {Bonitz},\ and\
  \citenamefont {Drescher}}]{2012Auger}%
  \BibitemOpen
  \bibfield  {author} {\bibinfo {author} {\bibfnamefont {B.}~\bibnamefont
  {Sch\"utte}}, \bibinfo {author} {\bibfnamefont {S.}~\bibnamefont {Bauch}},
  \bibinfo {author} {\bibfnamefont {U.}~\bibnamefont {Fr\"uhling}}, \bibinfo
  {author} {\bibfnamefont {M.}~\bibnamefont {Wieland}}, \bibinfo {author}
  {\bibfnamefont {M.}~\bibnamefont {Gensch}}, \bibinfo {author} {\bibfnamefont
  {E.}~\bibnamefont {Pl\"onjes}}, \bibinfo {author} {\bibfnamefont
  {T.}~\bibnamefont {Gaumnitz}}, \bibinfo {author} {\bibfnamefont
  {A.}~\bibnamefont {Azima}}, \bibinfo {author} {\bibfnamefont
  {M.}~\bibnamefont {Bonitz}}, \ and\ \bibinfo {author} {\bibfnamefont
  {M.}~\bibnamefont {Drescher}},\ }\href {\doibase
  10.1103/PhysRevLett.108.253003} {\bibfield  {journal} {\bibinfo  {journal}
  {Phys. Rev. Lett.}\ }\textbf {\bibinfo {volume} {108}},\ \bibinfo {pages}
  {253003} (\bibinfo {year} {2012})}\BibitemShut {NoStop}%
\bibitem [{\citenamefont {Remetter}\ \emph {et~al.}(2006)\citenamefont
  {Remetter}, \citenamefont {Johnsson}, \citenamefont {Mauritsson},
  \citenamefont {Varju}, \citenamefont {Ni}, \citenamefont {Lepine},
  \citenamefont {Gustafsson}, \citenamefont {Kling}, \citenamefont {Khan},
  \citenamefont {Lopez-Martens}, \citenamefont {Schafer}, \citenamefont
  {Vrakking},\ and\ \citenamefont {L'Huillier}}]{2006electroninterferometry}%
  \BibitemOpen
  \bibfield  {author} {\bibinfo {author} {\bibfnamefont {T.}~\bibnamefont
  {Remetter}}, \bibinfo {author} {\bibfnamefont {P.}~\bibnamefont {Johnsson}},
  \bibinfo {author} {\bibfnamefont {J.}~\bibnamefont {Mauritsson}}, \bibinfo
  {author} {\bibfnamefont {K.}~\bibnamefont {Varju}}, \bibinfo {author}
  {\bibfnamefont {Y.}~\bibnamefont {Ni}}, \bibinfo {author} {\bibfnamefont
  {F.}~\bibnamefont {Lepine}}, \bibinfo {author} {\bibfnamefont
  {E.}~\bibnamefont {Gustafsson}}, \bibinfo {author} {\bibfnamefont
  {M.}~\bibnamefont {Kling}}, \bibinfo {author} {\bibfnamefont
  {J.}~\bibnamefont {Khan}}, \bibinfo {author} {\bibfnamefont {R.}~\bibnamefont
  {Lopez-Martens}}, \bibinfo {author} {\bibfnamefont {K.}~\bibnamefont
  {Schafer}}, \bibinfo {author} {\bibfnamefont {M.}~\bibnamefont {Vrakking}}, \
  and\ \bibinfo {author} {\bibfnamefont {A.}~\bibnamefont {L'Huillier}},\
  }\href {\doibase 10.1038/nphys290} {\bibfield  {journal} {\bibinfo  {journal}
  {Nature Physics}\ }\textbf {\bibinfo {volume} {2}},\ \bibinfo {pages} {323}
  (\bibinfo {year} {2006})}\BibitemShut {NoStop}%
\bibitem [{\citenamefont {Goulielmakis}\ \emph {et~al.}(2010)\citenamefont
  {Goulielmakis}, \citenamefont {Loh}, \citenamefont {Wirth}, \citenamefont
  {Santra}, \citenamefont {Rohringer}, \citenamefont {Yakovlev}, \citenamefont
  {Zherebtsov}, \citenamefont {Pfeifer}, \citenamefont {Azzeer}, \citenamefont
  {Kling}, \citenamefont {Leone},\ and\ \citenamefont {Krausz}}]{2010realtime}%
  \BibitemOpen
  \bibfield  {author} {\bibinfo {author} {\bibfnamefont {E.}~\bibnamefont
  {Goulielmakis}}, \bibinfo {author} {\bibfnamefont {Z.-H.}\ \bibnamefont
  {Loh}}, \bibinfo {author} {\bibfnamefont {A.}~\bibnamefont {Wirth}}, \bibinfo
  {author} {\bibfnamefont {R.}~\bibnamefont {Santra}}, \bibinfo {author}
  {\bibfnamefont {N.}~\bibnamefont {Rohringer}}, \bibinfo {author}
  {\bibfnamefont {V.~S.}\ \bibnamefont {Yakovlev}}, \bibinfo {author}
  {\bibfnamefont {S.}~\bibnamefont {Zherebtsov}}, \bibinfo {author}
  {\bibfnamefont {T.}~\bibnamefont {Pfeifer}}, \bibinfo {author} {\bibfnamefont
  {A.~M.}\ \bibnamefont {Azzeer}}, \bibinfo {author} {\bibfnamefont {M.~F.}\
  \bibnamefont {Kling}}, \bibinfo {author} {\bibfnamefont {S.~R.}\ \bibnamefont
  {Leone}}, \ and\ \bibinfo {author} {\bibfnamefont {F.}~\bibnamefont
  {Krausz}},\ }\href {\doibase 10.1038/nature09212} {\bibfield  {journal}
  {\bibinfo  {journal} {Nature (London)}\ }\textbf {\bibinfo {volume} {466}},\
  \bibinfo {pages} {739} (\bibinfo {year} {2010})}\BibitemShut {NoStop}%
\bibitem [{\citenamefont {Wirth}\ \emph {et~al.}(2011)\citenamefont {Wirth},
  \citenamefont {Hassan}, \citenamefont {Grguras}, \citenamefont {Gagnon},
  \citenamefont {Moulet}, \citenamefont {Luu}, \citenamefont {Pabst},
  \citenamefont {Santra}, \citenamefont {Alahmed}, \citenamefont {Azzeer},
  \citenamefont {Yakovlev}, \citenamefont {Pervak}, \citenamefont {Krausz},\
  and\ \citenamefont {Goulielmakis}}]{Wirth2011}%
  \BibitemOpen
  \bibfield  {author} {\bibinfo {author} {\bibfnamefont {A.}~\bibnamefont
  {Wirth}}, \bibinfo {author} {\bibfnamefont {M.~T.}\ \bibnamefont {Hassan}},
  \bibinfo {author} {\bibfnamefont {I.}~\bibnamefont {Grguras}}, \bibinfo
  {author} {\bibfnamefont {J.}~\bibnamefont {Gagnon}}, \bibinfo {author}
  {\bibfnamefont {A.}~\bibnamefont {Moulet}}, \bibinfo {author} {\bibfnamefont
  {T.~T.}\ \bibnamefont {Luu}}, \bibinfo {author} {\bibfnamefont
  {S.}~\bibnamefont {Pabst}}, \bibinfo {author} {\bibfnamefont
  {R.}~\bibnamefont {Santra}}, \bibinfo {author} {\bibfnamefont {Z.~A.}\
  \bibnamefont {Alahmed}}, \bibinfo {author} {\bibfnamefont {A.~M.}\
  \bibnamefont {Azzeer}}, \bibinfo {author} {\bibfnamefont {V.~S.}\
  \bibnamefont {Yakovlev}}, \bibinfo {author} {\bibfnamefont {V.}~\bibnamefont
  {Pervak}}, \bibinfo {author} {\bibfnamefont {F.}~\bibnamefont {Krausz}}, \
  and\ \bibinfo {author} {\bibfnamefont {E.}~\bibnamefont {Goulielmakis}},\
  }\href {\doibase 10.1126/science.1210268} {\bibfield  {journal} {\bibinfo
  {journal} {Science}\ }\textbf {\bibinfo {volume} {334}},\ \bibinfo {pages}
  {195} (\bibinfo {year} {2011})}\BibitemShut {NoStop}%
\bibitem [{\citenamefont {Huang}\ \emph {et~al.}(2012)\citenamefont {Huang},
  \citenamefont {Cirmi}, \citenamefont {Moses}, \citenamefont {Hong},
  \citenamefont {Bhardwaj}, \citenamefont {Birge}, \citenamefont {Chen},
  \citenamefont {Kabakova}, \citenamefont {Li}, \citenamefont {Eggleton},
  \citenamefont {Cerullo},\ and\ \citenamefont {Kartner}}]{2012LFS}%
  \BibitemOpen
  \bibfield  {author} {\bibinfo {author} {\bibfnamefont {S.-W.}\ \bibnamefont
  {Huang}}, \bibinfo {author} {\bibfnamefont {G.}~\bibnamefont {Cirmi}},
  \bibinfo {author} {\bibfnamefont {J.}~\bibnamefont {Moses}}, \bibinfo
  {author} {\bibfnamefont {K.-H.}\ \bibnamefont {Hong}}, \bibinfo {author}
  {\bibfnamefont {S.}~\bibnamefont {Bhardwaj}}, \bibinfo {author}
  {\bibfnamefont {J.~R.}\ \bibnamefont {Birge}}, \bibinfo {author}
  {\bibfnamefont {L.-J.}\ \bibnamefont {Chen}}, \bibinfo {author}
  {\bibfnamefont {I.~V.}\ \bibnamefont {Kabakova}}, \bibinfo {author}
  {\bibfnamefont {E.}~\bibnamefont {Li}}, \bibinfo {author} {\bibfnamefont
  {B.~J.}\ \bibnamefont {Eggleton}}, \bibinfo {author} {\bibfnamefont
  {G.}~\bibnamefont {Cerullo}}, \ and\ \bibinfo {author} {\bibfnamefont
  {F.~X.}\ \bibnamefont {Kartner}},\ }\href
  {http://stacks.iop.org/0953-4075/45/i=7/a=074009} {\bibfield  {journal}
  {\bibinfo  {journal} {Journal of Physics B}\ }\textbf {\bibinfo {volume}
  {45}},\ \bibinfo {pages} {074009} (\bibinfo {year} {2012})}\BibitemShut
  {NoStop}%
\bibitem [{\citenamefont {Goulielmakis}\ \emph {et~al.}(2008)\citenamefont
  {Goulielmakis}, \citenamefont {Schultze}, \citenamefont {Hofstetter},
  \citenamefont {Yakovlev}, \citenamefont {Gagnon}, \citenamefont {Uiberacker},
  \citenamefont {Aquila}, \citenamefont {Gullikson}, \citenamefont {Attwood},
  \citenamefont {Kienberger}, \citenamefont {Krausz},\ and\ \citenamefont
  {Kleineberg}}]{80as}%
  \BibitemOpen
  \bibfield  {author} {\bibinfo {author} {\bibfnamefont {E.}~\bibnamefont
  {Goulielmakis}}, \bibinfo {author} {\bibfnamefont {M.}~\bibnamefont
  {Schultze}}, \bibinfo {author} {\bibfnamefont {M.}~\bibnamefont
  {Hofstetter}}, \bibinfo {author} {\bibfnamefont {V.~S.}\ \bibnamefont
  {Yakovlev}}, \bibinfo {author} {\bibfnamefont {J.}~\bibnamefont {Gagnon}},
  \bibinfo {author} {\bibfnamefont {M.}~\bibnamefont {Uiberacker}}, \bibinfo
  {author} {\bibfnamefont {A.~L.}\ \bibnamefont {Aquila}}, \bibinfo {author}
  {\bibfnamefont {E.~M.}\ \bibnamefont {Gullikson}}, \bibinfo {author}
  {\bibfnamefont {D.~T.}\ \bibnamefont {Attwood}}, \bibinfo {author}
  {\bibfnamefont {R.}~\bibnamefont {Kienberger}}, \bibinfo {author}
  {\bibfnamefont {F.}~\bibnamefont {Krausz}}, \ and\ \bibinfo {author}
  {\bibfnamefont {U.}~\bibnamefont {Kleineberg}},\ }\href {\doibase
  10.1126/science.1157846} {\bibfield  {journal} {\bibinfo  {journal}
  {Science}\ }\textbf {\bibinfo {volume} {320}},\ \bibinfo {pages} {1614}
  (\bibinfo {year} {2008})}\BibitemShut {NoStop}%
\bibitem [{\citenamefont {Zhao}\ \emph {et~al.}(2012)\citenamefont {Zhao},
  \citenamefont {Zhang}, \citenamefont {Chini}, \citenamefont {Wu},
  \citenamefont {Wang},\ and\ \citenamefont {Chang}}]{Zhao2012}%
  \BibitemOpen
  \bibfield  {author} {\bibinfo {author} {\bibfnamefont {K.}~\bibnamefont
  {Zhao}}, \bibinfo {author} {\bibfnamefont {Q.}~\bibnamefont {Zhang}},
  \bibinfo {author} {\bibfnamefont {M.}~\bibnamefont {Chini}}, \bibinfo
  {author} {\bibfnamefont {Y.}~\bibnamefont {Wu}}, \bibinfo {author}
  {\bibfnamefont {X.}~\bibnamefont {Wang}}, \ and\ \bibinfo {author}
  {\bibfnamefont {Z.}~\bibnamefont {Chang}},\ }\href {\doibase
  10.1364/OL.37.003891} {\bibfield  {journal} {\bibinfo  {journal} {Opt.
  Lett.}\ }\textbf {\bibinfo {volume} {37}},\ \bibinfo {pages} {3891} (\bibinfo
  {year} {2012})}\BibitemShut {NoStop}%
\bibitem [{\citenamefont {Tzallas}\ \emph {et~al.}(2011)\citenamefont
  {Tzallas}, \citenamefont {Skantzakis}, \citenamefont {Nikolopoulos},
  \citenamefont {Tsakiris},\ and\ \citenamefont {Charalambidis}}]{2011xuvpp}%
  \BibitemOpen
  \bibfield  {author} {\bibinfo {author} {\bibfnamefont {P.}~\bibnamefont
  {Tzallas}}, \bibinfo {author} {\bibfnamefont {E.}~\bibnamefont {Skantzakis}},
  \bibinfo {author} {\bibfnamefont {L.~A.~A.}\ \bibnamefont {Nikolopoulos}},
  \bibinfo {author} {\bibfnamefont {G.~D.}\ \bibnamefont {Tsakiris}}, \ and\
  \bibinfo {author} {\bibfnamefont {D.}~\bibnamefont {Charalambidis}},\ }\href
  {\doibase 10.1038/nphys2033} {\bibfield  {journal} {\bibinfo  {journal}
  {Nature Physics}\ }\textbf {\bibinfo {volume} {7}},\ \bibinfo {pages} {781}
  (\bibinfo {year} {2011})}\BibitemShut {NoStop}%
\bibitem [{\citenamefont {Popmintchev}\ \emph {et~al.}(2010)\citenamefont
  {Popmintchev}, \citenamefont {Chen}, \citenamefont {Arpin}, \citenamefont
  {Murnane},\ and\ \citenamefont {Kapteyn}}]{2010NPhotPopmintchev}%
  \BibitemOpen
  \bibfield  {author} {\bibinfo {author} {\bibfnamefont {T.}~\bibnamefont
  {Popmintchev}}, \bibinfo {author} {\bibfnamefont {M.-C.}\ \bibnamefont
  {Chen}}, \bibinfo {author} {\bibfnamefont {P.}~\bibnamefont {Arpin}},
  \bibinfo {author} {\bibfnamefont {M.~M.}\ \bibnamefont {Murnane}}, \ and\
  \bibinfo {author} {\bibfnamefont {H.~C.}\ \bibnamefont {Kapteyn}},\ }\href
  {\doibase 10.1038/nphoton.2010.256} {\bibfield  {journal} {\bibinfo
  {journal} {Nature Photonics}\ }\textbf {\bibinfo {volume} {4}},\ \bibinfo
  {pages} {822} (\bibinfo {year} {2010})}\BibitemShut {NoStop}%
\bibitem [{\citenamefont {Popmintchev}\ \emph {et~al.}(2012)\citenamefont
  {Popmintchev}, \citenamefont {Chen}, \citenamefont {Popmintchev},
  \citenamefont {Arpin}, \citenamefont {Brown}, \citenamefont {Alisauskas},
  \citenamefont {Andriukaitis}, \citenamefont {Balciunas}, \citenamefont
  {M\"ucke}, \citenamefont {Pugzlys}, \citenamefont {Baltuska}, \citenamefont
  {Shim}, \citenamefont {Schrauth}, \citenamefont {Gaeta}, \citenamefont
  {Hernández-García}, \citenamefont {Plaja}, \citenamefont {Becker},
  \citenamefont {Jaron-Becker}, \citenamefont {Murnane},\ and\ \citenamefont
  {Kapteyn}}]{2012kev}%
  \BibitemOpen
  \bibfield  {author} {\bibinfo {author} {\bibfnamefont {T.}~\bibnamefont
  {Popmintchev}}, \bibinfo {author} {\bibfnamefont {M.-C.}\ \bibnamefont
  {Chen}}, \bibinfo {author} {\bibfnamefont {D.}~\bibnamefont {Popmintchev}},
  \bibinfo {author} {\bibfnamefont {P.}~\bibnamefont {Arpin}}, \bibinfo
  {author} {\bibfnamefont {S.}~\bibnamefont {Brown}}, \bibinfo {author}
  {\bibfnamefont {S.}~\bibnamefont {Alisauskas}}, \bibinfo {author}
  {\bibfnamefont {G.}~\bibnamefont {Andriukaitis}}, \bibinfo {author}
  {\bibfnamefont {T.}~\bibnamefont {Balciunas}}, \bibinfo {author}
  {\bibfnamefont {O.~D.}\ \bibnamefont {M\"ucke}}, \bibinfo {author}
  {\bibfnamefont {A.}~\bibnamefont {Pugzlys}}, \bibinfo {author} {\bibfnamefont
  {A.}~\bibnamefont {Baltuska}}, \bibinfo {author} {\bibfnamefont
  {B.}~\bibnamefont {Shim}}, \bibinfo {author} {\bibfnamefont {S.~E.}\
  \bibnamefont {Schrauth}}, \bibinfo {author} {\bibfnamefont {A.}~\bibnamefont
  {Gaeta}}, \bibinfo {author} {\bibfnamefont {C.}~\bibnamefont
  {Hernández-García}}, \bibinfo {author} {\bibfnamefont {L.}~\bibnamefont
  {Plaja}}, \bibinfo {author} {\bibfnamefont {A.}~\bibnamefont {Becker}},
  \bibinfo {author} {\bibfnamefont {A.}~\bibnamefont {Jaron-Becker}}, \bibinfo
  {author} {\bibfnamefont {M.~M.}\ \bibnamefont {Murnane}}, \ and\ \bibinfo
  {author} {\bibfnamefont {H.~C.}\ \bibnamefont {Kapteyn}},\ }\href {\doibase
  10.1126/science.1218497} {\bibfield  {journal} {\bibinfo  {journal}
  {Science}\ }\textbf {\bibinfo {volume} {336}},\ \bibinfo {pages} {1287}
  (\bibinfo {year} {2012})}\BibitemShut {NoStop}%
\bibitem [{\citenamefont {Sansone}\ \emph {et~al.}(2011)\citenamefont
  {Sansone}, \citenamefont {Poletto},\ and\ \citenamefont
  {Nisoli}}]{2011Sansone}%
  \BibitemOpen
  \bibfield  {author} {\bibinfo {author} {\bibfnamefont {G.}~\bibnamefont
  {Sansone}}, \bibinfo {author} {\bibfnamefont {L.}~\bibnamefont {Poletto}}, \
  and\ \bibinfo {author} {\bibfnamefont {M.}~\bibnamefont {Nisoli}},\ }\href
  {\doibase 10.1038/NPHOTON.2011.167} {\bibfield  {journal} {\bibinfo
  {journal} {Nature Photonics}\ }\textbf {\bibinfo {volume} {5}},\ \bibinfo
  {pages} {656} (\bibinfo {year} {2011})}\BibitemShut {NoStop}%
\bibitem [{\citenamefont {Takahashi}\ \emph {et~al.}(2013)\citenamefont
  {Takahashi}, \citenamefont {Lan}, \citenamefont {M\"ucke}, \citenamefont
  {Nabekawa},\ and\ \citenamefont {Midorikawa}}]{2013nctakahashi}%
  \BibitemOpen
  \bibfield  {author} {\bibinfo {author} {\bibfnamefont {E.~J.}\ \bibnamefont
  {Takahashi}}, \bibinfo {author} {\bibfnamefont {P.}~\bibnamefont {Lan}},
  \bibinfo {author} {\bibfnamefont {O.~D.}\ \bibnamefont {M\"ucke}}, \bibinfo
  {author} {\bibfnamefont {Y.}~\bibnamefont {Nabekawa}}, \ and\ \bibinfo
  {author} {\bibfnamefont {K.}~\bibnamefont {Midorikawa}},\ }\href {\doibase
  10.1038/ncomms3691} {\bibfield  {journal} {\bibinfo  {journal} {Nature
  Communications}\ }\textbf {\bibinfo {volume} {4}},\ \bibinfo {pages} {2691}
  (\bibinfo {year} {2013})}\BibitemShut {NoStop}%
\bibitem [{\citenamefont {Heyl}\ \emph {et~al.}(2012)\citenamefont {Heyl},
  \citenamefont {Gudde}, \citenamefont {L'Huillier},\ and\ \citenamefont
  {Hofer}}]{2012khz}%
  \BibitemOpen
  \bibfield  {author} {\bibinfo {author} {\bibfnamefont {C.~M.}\ \bibnamefont
  {Heyl}}, \bibinfo {author} {\bibfnamefont {J.}~\bibnamefont {Gudde}},
  \bibinfo {author} {\bibfnamefont {A.}~\bibnamefont {L'Huillier}}, \ and\
  \bibinfo {author} {\bibfnamefont {U.}~\bibnamefont {Hofer}},\ }\href
  {http://stacks.iop.org/0953-4075/45/i=7/a=074020} {\bibfield  {journal}
  {\bibinfo  {journal} {Journal of Physics B}\ }\textbf {\bibinfo {volume}
  {45}},\ \bibinfo {pages} {074020} (\bibinfo {year} {2012})}\BibitemShut
  {NoStop}%
\bibitem [{\citenamefont {Mills}\ \emph {et~al.}(2012)\citenamefont {Mills},
  \citenamefont {Hammond}, \citenamefont {Lam},\ and\ \citenamefont
  {Jones}}]{2012incavity}%
  \BibitemOpen
  \bibfield  {author} {\bibinfo {author} {\bibfnamefont {A.~K.}\ \bibnamefont
  {Mills}}, \bibinfo {author} {\bibfnamefont {T.~J.}\ \bibnamefont {Hammond}},
  \bibinfo {author} {\bibfnamefont {M.~H.~C.}\ \bibnamefont {Lam}}, \ and\
  \bibinfo {author} {\bibfnamefont {D.~J.}\ \bibnamefont {Jones}},\ }\href
  {http://stacks.iop.org/0953-4075/45/i=14/a=142001} {\bibfield  {journal}
  {\bibinfo  {journal} {Journal of Physics B}\ }\textbf {\bibinfo {volume}
  {45}},\ \bibinfo {pages} {142001} (\bibinfo {year} {2012})}\BibitemShut
  {NoStop}%
\bibitem [{\citenamefont {Krebs}\ \emph {et~al.}(2013)\citenamefont {Krebs},
  \citenamefont {Hadrich}, \citenamefont {Demmler}, \citenamefont {Rothhardt},
  \citenamefont {Zair}, \citenamefont {Chipperfield}, \citenamefont {Limpert},\
  and\ \citenamefont {Tunnermann}}]{2013mhzhhg}%
  \BibitemOpen
  \bibfield  {author} {\bibinfo {author} {\bibfnamefont {M.}~\bibnamefont
  {Krebs}}, \bibinfo {author} {\bibfnamefont {S.}~\bibnamefont {Hadrich}},
  \bibinfo {author} {\bibfnamefont {S.}~\bibnamefont {Demmler}}, \bibinfo
  {author} {\bibfnamefont {J.}~\bibnamefont {Rothhardt}}, \bibinfo {author}
  {\bibfnamefont {A.}~\bibnamefont {Zair}}, \bibinfo {author} {\bibfnamefont
  {L.}~\bibnamefont {Chipperfield}}, \bibinfo {author} {\bibfnamefont
  {J.}~\bibnamefont {Limpert}}, \ and\ \bibinfo {author} {\bibfnamefont
  {A.}~\bibnamefont {Tunnermann}},\ }\href {\doibase 10.1038/NPHOTON.2013.131}
  {\bibfield  {journal} {\bibinfo  {journal} {Nature Photonics}\ }\textbf
  {\bibinfo {volume} {7}},\ \bibinfo {pages} {555} (\bibinfo {year}
  {2013})}\BibitemShut {NoStop}%
\bibitem [{\citenamefont {Roos}\ \emph {et~al.}(2001)\citenamefont {Roos},
  \citenamefont {Gaarde},\ and\ \citenamefont {L'Huillier}}]{2001roos}%
  \BibitemOpen
  \bibfield  {author} {\bibinfo {author} {\bibfnamefont {L.}~\bibnamefont
  {Roos}}, \bibinfo {author} {\bibfnamefont {M.~B.}\ \bibnamefont {Gaarde}}, \
  and\ \bibinfo {author} {\bibfnamefont {A.}~\bibnamefont {L'Huillier}},\
  }\href {http://stacks.iop.org/0953-4075/34/i=24/a=307} {\bibfield  {journal}
  {\bibinfo  {journal} {Journal of Physics B}\ }\textbf {\bibinfo {volume}
  {34}},\ \bibinfo {pages} {5041} (\bibinfo {year} {2001})}\BibitemShut
  {NoStop}%
\bibitem [{\citenamefont {Chu}\ and\ \citenamefont {Chu}(2001)}]{2001hhgoptim}%
  \BibitemOpen
  \bibfield  {author} {\bibinfo {author} {\bibfnamefont {X.}~\bibnamefont
  {Chu}}\ and\ \bibinfo {author} {\bibfnamefont {S.-I.}\ \bibnamefont {Chu}},\
  }\href {\doibase 10.1103/PhysRevA.64.021403} {\bibfield  {journal} {\bibinfo
  {journal} {Phys. Rev. A}\ }\textbf {\bibinfo {volume} {64}},\ \bibinfo
  {pages} {021403} (\bibinfo {year} {2001})}\BibitemShut {NoStop}%
\bibitem [{\citenamefont {Schaefer}\ and\ \citenamefont
  {Kosloff}(2012)}]{2012OCT}%
  \BibitemOpen
  \bibfield  {author} {\bibinfo {author} {\bibfnamefont {I.}~\bibnamefont
  {Schaefer}}\ and\ \bibinfo {author} {\bibfnamefont {R.}~\bibnamefont
  {Kosloff}},\ }\href {\doibase 10.1103/PhysRevA.86.063417} {\bibfield
  {journal} {\bibinfo  {journal} {Phys. Rev. A}\ }\textbf {\bibinfo {volume}
  {86}},\ \bibinfo {pages} {063417} (\bibinfo {year} {2012})}\BibitemShut
  {NoStop}%
\bibitem [{\citenamefont {Christov}\ \emph {et~al.}(2001)\citenamefont
  {Christov}, \citenamefont {Bartels}, \citenamefont {Kapteyn},\ and\
  \citenamefont {Murnane}}]{2001PRLChristov}%
  \BibitemOpen
  \bibfield  {author} {\bibinfo {author} {\bibfnamefont {I.~P.}\ \bibnamefont
  {Christov}}, \bibinfo {author} {\bibfnamefont {R.}~\bibnamefont {Bartels}},
  \bibinfo {author} {\bibfnamefont {H.~C.}\ \bibnamefont {Kapteyn}}, \ and\
  \bibinfo {author} {\bibfnamefont {M.~M.}\ \bibnamefont {Murnane}},\ }\href
  {\doibase 10.1103/PhysRevLett.86.5458} {\bibfield  {journal} {\bibinfo
  {journal} {Phys. Rev. Lett.}\ }\textbf {\bibinfo {volume} {86}},\ \bibinfo
  {pages} {5458} (\bibinfo {year} {2001})}\BibitemShut {NoStop}%
\bibitem [{\citenamefont {Ivanov}\ and\ \citenamefont
  {Kheifets}(2009)}]{2009cutoff}%
  \BibitemOpen
  \bibfield  {author} {\bibinfo {author} {\bibfnamefont {I.~A.}\ \bibnamefont
  {Ivanov}}\ and\ \bibinfo {author} {\bibfnamefont {A.~S.}\ \bibnamefont
  {Kheifets}},\ }\href {\doibase 10.1103/PhysRevA.80.023809} {\bibfield
  {journal} {\bibinfo  {journal} {Phys. Rev. A}\ }\textbf {\bibinfo {volume}
  {80}},\ \bibinfo {pages} {023809} (\bibinfo {year} {2009})}\BibitemShut
  {NoStop}%
\bibitem [{\citenamefont {Di}\ and\ \citenamefont
  {Fu-Li}(2013)}]{2013cutoffext}%
  \BibitemOpen
  \bibfield  {author} {\bibinfo {author} {\bibfnamefont {Z.}~\bibnamefont
  {Di}}\ and\ \bibinfo {author} {\bibfnamefont {L.}~\bibnamefont {Fu-Li}},\
  }\href {\doibase 10.1088/1674-1056/22/6/064215} {\bibfield  {journal}
  {\bibinfo  {journal} {Chinese Physics B}\ }\textbf {\bibinfo {volume} {22}},\
  \bibinfo {pages} {064215} (\bibinfo {year} {2013})}\BibitemShut {NoStop}%
\bibitem [{\citenamefont {Ben Haj~Yedder}\ \emph {et~al.}(2004)\citenamefont
  {Ben Haj~Yedder}, \citenamefont {Le~Bris}, \citenamefont {Atabek},
  \citenamefont {Chelkowski},\ and\ \citenamefont
  {Bandrauk}}]{2004sapoptimization}%
  \BibitemOpen
  \bibfield  {author} {\bibinfo {author} {\bibfnamefont {A.}~\bibnamefont {Ben
  Haj~Yedder}}, \bibinfo {author} {\bibfnamefont {C.}~\bibnamefont {Le~Bris}},
  \bibinfo {author} {\bibfnamefont {O.}~\bibnamefont {Atabek}}, \bibinfo
  {author} {\bibfnamefont {S.}~\bibnamefont {Chelkowski}}, \ and\ \bibinfo
  {author} {\bibfnamefont {A.~D.}\ \bibnamefont {Bandrauk}},\ }\href {\doibase
  10.1103/PhysRevA.69.041802} {\bibfield  {journal} {\bibinfo  {journal} {Phys.
  Rev. A}\ }\textbf {\bibinfo {volume} {69}},\ \bibinfo {pages} {041802}
  (\bibinfo {year} {2004})}\BibitemShut {NoStop}%
\bibitem [{\citenamefont {Tang}\ \emph {et~al.}(2010)\citenamefont {Tang},
  \citenamefont {Zheng},\ and\ \citenamefont {Chen}}]{Tang2010155}%
  \BibitemOpen
  \bibfield  {author} {\bibinfo {author} {\bibfnamefont {S.}~\bibnamefont
  {Tang}}, \bibinfo {author} {\bibfnamefont {L.}~\bibnamefont {Zheng}}, \ and\
  \bibinfo {author} {\bibfnamefont {X.}~\bibnamefont {Chen}},\ }\href {\doibase
  10.1016/j.optcom.2009.09.076} {\bibfield  {journal} {\bibinfo  {journal}
  {Optics Communications}\ }\textbf {\bibinfo {volume} {283}},\ \bibinfo
  {pages} {155 } (\bibinfo {year} {2010})}\BibitemShut {NoStop}%
\bibitem [{\citenamefont {Bartels}\ \emph {et~al.}(2000)\citenamefont
  {Bartels}, \citenamefont {Backus}, \citenamefont {Zeek}, \citenamefont
  {Misoguti}, \citenamefont {Vdovin}, \citenamefont {Christov}, \citenamefont
  {Murnane},\ and\ \citenamefont {Kapteyn}}]{2000Bartels}%
  \BibitemOpen
  \bibfield  {author} {\bibinfo {author} {\bibfnamefont {R.}~\bibnamefont
  {Bartels}}, \bibinfo {author} {\bibfnamefont {S.}~\bibnamefont {Backus}},
  \bibinfo {author} {\bibfnamefont {E.}~\bibnamefont {Zeek}}, \bibinfo {author}
  {\bibfnamefont {L.}~\bibnamefont {Misoguti}}, \bibinfo {author}
  {\bibfnamefont {G.}~\bibnamefont {Vdovin}}, \bibinfo {author} {\bibfnamefont
  {I.}~\bibnamefont {Christov}}, \bibinfo {author} {\bibfnamefont
  {M.}~\bibnamefont {Murnane}}, \ and\ \bibinfo {author} {\bibfnamefont
  {H.}~\bibnamefont {Kapteyn}},\ }\href {\doibase 10.1038/35018029} {\bibfield
  {journal} {\bibinfo  {journal} {Nature (London)}\ }\textbf {\bibinfo {volume}
  {406}},\ \bibinfo {pages} {164} (\bibinfo {year} {2000})}\BibitemShut
  {NoStop}%
\bibitem [{\citenamefont {Bartels}\ \emph {et~al.}(2004)\citenamefont
  {Bartels}, \citenamefont {Murnane}, \citenamefont {Kapteyn}, \citenamefont
  {Christov},\ and\ \citenamefont {Rabitz}}]{2004Bartels}%
  \BibitemOpen
  \bibfield  {author} {\bibinfo {author} {\bibfnamefont {R.~A.}\ \bibnamefont
  {Bartels}}, \bibinfo {author} {\bibfnamefont {M.~M.}\ \bibnamefont
  {Murnane}}, \bibinfo {author} {\bibfnamefont {H.~C.}\ \bibnamefont
  {Kapteyn}}, \bibinfo {author} {\bibfnamefont {I.}~\bibnamefont {Christov}}, \
  and\ \bibinfo {author} {\bibfnamefont {H.}~\bibnamefont {Rabitz}},\ }\href
  {\doibase 10.1103/PhysRevA.70.043404} {\bibfield  {journal} {\bibinfo
  {journal} {Phys. Rev. A}\ }\textbf {\bibinfo {volume} {70}},\ \bibinfo
  {pages} {043404} (\bibinfo {year} {2004})}\BibitemShut {NoStop}%
\bibitem [{\citenamefont {Winterfeldt}\ \emph {et~al.}(2008)\citenamefont
  {Winterfeldt}, \citenamefont {Spielmann},\ and\ \citenamefont
  {Gerber}}]{2008RevModPhys}%
  \BibitemOpen
  \bibfield  {author} {\bibinfo {author} {\bibfnamefont {C.}~\bibnamefont
  {Winterfeldt}}, \bibinfo {author} {\bibfnamefont {C.}~\bibnamefont
  {Spielmann}}, \ and\ \bibinfo {author} {\bibfnamefont {G.}~\bibnamefont
  {Gerber}},\ }\href {\doibase 10.1103/RevModPhys.80.117} {\bibfield  {journal}
  {\bibinfo  {journal} {Rev. Mod. Phys.}\ }\textbf {\bibinfo {volume} {80}},\
  \bibinfo {pages} {117} (\bibinfo {year} {2008})}\BibitemShut {NoStop}%
\bibitem [{\citenamefont {Spitzenpfeil}\ \emph {et~al.}(2009)\citenamefont
  {Spitzenpfeil}, \citenamefont {Eyring}, \citenamefont {Kern}, \citenamefont
  {Ott}, \citenamefont {Lohbreier}, \citenamefont {Henneberger}, \citenamefont
  {Franke}, \citenamefont {Jung}, \citenamefont {Walter}, \citenamefont
  {Weger}, \citenamefont {Winterfeldt}, \citenamefont {Pfeifer},\ and\
  \citenamefont {Spielmann}}]{raey}%
  \BibitemOpen
  \bibfield  {author} {\bibinfo {author} {\bibfnamefont {R.}~\bibnamefont
  {Spitzenpfeil}}, \bibinfo {author} {\bibfnamefont {S.}~\bibnamefont
  {Eyring}}, \bibinfo {author} {\bibfnamefont {C.}~\bibnamefont {Kern}},
  \bibinfo {author} {\bibfnamefont {C.}~\bibnamefont {Ott}}, \bibinfo {author}
  {\bibfnamefont {J.}~\bibnamefont {Lohbreier}}, \bibinfo {author}
  {\bibfnamefont {J.}~\bibnamefont {Henneberger}}, \bibinfo {author}
  {\bibfnamefont {N.}~\bibnamefont {Franke}}, \bibinfo {author} {\bibfnamefont
  {S.}~\bibnamefont {Jung}}, \bibinfo {author} {\bibfnamefont {D.}~\bibnamefont
  {Walter}}, \bibinfo {author} {\bibfnamefont {M.}~\bibnamefont {Weger}},
  \bibinfo {author} {\bibfnamefont {C.}~\bibnamefont {Winterfeldt}}, \bibinfo
  {author} {\bibfnamefont {T.}~\bibnamefont {Pfeifer}}, \ and\ \bibinfo
  {author} {\bibfnamefont {C.}~\bibnamefont {Spielmann}},\ }\href {\doibase
  10.1007/s00339-009-5173-7} {\bibfield  {journal} {\bibinfo  {journal}
  {Applied Physics A}\ }\textbf {\bibinfo {volume} {96}},\ \bibinfo {pages}
  {69} (\bibinfo {year} {2009})}\BibitemShut {NoStop}%
\bibitem [{\citenamefont {et~al.}(2014)}]{lefteris}%
  \BibitemOpen
  \bibfield  {author} {\bibinfo {author} {\bibfnamefont {H.~M.~T.}\
  \bibnamefont {et~al.}},\ }\href@noop {} {\bibfield  {journal} {\bibinfo
  {journal} {submitted}\ } (\bibinfo {year} {2014})}\BibitemShut {NoStop}%
\bibitem [{\citenamefont {Lewenstein}\ \emph {et~al.}(1994)\citenamefont
  {Lewenstein}, \citenamefont {Balcou}, \citenamefont {Ivanov}, \citenamefont
  {L'Huillier},\ and\ \citenamefont {Corkum}}]{lewenstein}%
  \BibitemOpen
  \bibfield  {author} {\bibinfo {author} {\bibfnamefont {M.}~\bibnamefont
  {Lewenstein}}, \bibinfo {author} {\bibfnamefont {P.}~\bibnamefont {Balcou}},
  \bibinfo {author} {\bibfnamefont {M.~Y.}\ \bibnamefont {Ivanov}}, \bibinfo
  {author} {\bibfnamefont {A.}~\bibnamefont {L'Huillier}}, \ and\ \bibinfo
  {author} {\bibfnamefont {P.~B.}\ \bibnamefont {Corkum}},\ }\href {\doibase
  10.1103/PhysRevA.49.2117} {\bibfield  {journal} {\bibinfo  {journal} {Phys.
  Rev. A}\ }\textbf {\bibinfo {volume} {49}},\ \bibinfo {pages} {2117}
  (\bibinfo {year} {1994})}\BibitemShut {NoStop}%
\bibitem [{\citenamefont {Lewenstein}\ \emph {et~al.}(1995)\citenamefont
  {Lewenstein}, \citenamefont {Sali\`eres},\ and\ \citenamefont
  {L'Huillier}}]{lewph}%
  \BibitemOpen
  \bibfield  {author} {\bibinfo {author} {\bibfnamefont {M.}~\bibnamefont
  {Lewenstein}}, \bibinfo {author} {\bibfnamefont {P.}~\bibnamefont
  {Sali\`eres}}, \ and\ \bibinfo {author} {\bibfnamefont {A.}~\bibnamefont
  {L'Huillier}},\ }\href {\doibase 10.1103/PhysRevA.52.4747} {\bibfield
  {journal} {\bibinfo  {journal} {Phys. Rev. A}\ }\textbf {\bibinfo {volume}
  {52}},\ \bibinfo {pages} {4747} (\bibinfo {year} {1995})}\BibitemShut
  {NoStop}%
\bibitem [{\citenamefont {Lopez-Martens}\ \emph {et~al.}(2005)\citenamefont
  {Lopez-Martens}, \citenamefont {Varju}, \citenamefont {Johnsson},
  \citenamefont {Mauritsson}, \citenamefont {Mairesse}, \citenamefont
  {Sali\`eres}, \citenamefont {Gaarde}, \citenamefont {Schafer}, \citenamefont
  {Persson}, \citenamefont {Svanberg}, \citenamefont {Wahlstrom},\ and\
  \citenamefont {L'Huillier}}]{lopezprl}%
  \BibitemOpen
  \bibfield  {author} {\bibinfo {author} {\bibfnamefont {R.}~\bibnamefont
  {Lopez-Martens}}, \bibinfo {author} {\bibfnamefont {K.}~\bibnamefont
  {Varju}}, \bibinfo {author} {\bibfnamefont {P.}~\bibnamefont {Johnsson}},
  \bibinfo {author} {\bibfnamefont {J.}~\bibnamefont {Mauritsson}}, \bibinfo
  {author} {\bibfnamefont {Y.}~\bibnamefont {Mairesse}}, \bibinfo {author}
  {\bibfnamefont {P.}~\bibnamefont {Sali\`eres}}, \bibinfo {author}
  {\bibfnamefont {M.}~\bibnamefont {Gaarde}}, \bibinfo {author} {\bibfnamefont
  {K.}~\bibnamefont {Schafer}}, \bibinfo {author} {\bibfnamefont
  {A.}~\bibnamefont {Persson}}, \bibinfo {author} {\bibfnamefont
  {S.}~\bibnamefont {Svanberg}}, \bibinfo {author} {\bibfnamefont
  {C.}~\bibnamefont {Wahlstrom}}, \ and\ \bibinfo {author} {\bibfnamefont
  {A.}~\bibnamefont {L'Huillier}},\ }\href {\doibase
  10.1103/PhysRevLett.94.033001} {\bibfield  {journal} {\bibinfo  {journal}
  {Phys. Rev. Lett.}\ }\textbf {\bibinfo {volume} {94}},\ \bibinfo {pages}
  {033001} (\bibinfo {year} {2005})}\BibitemShut {NoStop}%
\bibitem [{\citenamefont {Takahashi}\ \emph {et~al.}(2003)\citenamefont
  {Takahashi}, \citenamefont {Tosa}, \citenamefont {Nabekawa},\ and\
  \citenamefont {Midorikawa}}]{tosamodel}%
  \BibitemOpen
  \bibfield  {author} {\bibinfo {author} {\bibfnamefont {E.}~\bibnamefont
  {Takahashi}}, \bibinfo {author} {\bibfnamefont {V.}~\bibnamefont {Tosa}},
  \bibinfo {author} {\bibfnamefont {Y.}~\bibnamefont {Nabekawa}}, \ and\
  \bibinfo {author} {\bibfnamefont {K.}~\bibnamefont {Midorikawa}},\ }\href
  {\doibase 10.1103/PhysRevA.68.023808} {\bibfield  {journal} {\bibinfo
  {journal} {Phys. Rev. A}\ }\textbf {\bibinfo {volume} {68}},\ \bibinfo
  {pages} {023808} (\bibinfo {year} {2003})}\BibitemShut {NoStop}%
\bibitem [{\citenamefont {Ammosov}\ \emph {et~al.}(1986)\citenamefont
  {Ammosov}, \citenamefont {Delone},\ and\ \citenamefont {Krainov}}]{adk}%
  \BibitemOpen
  \bibfield  {author} {\bibinfo {author} {\bibfnamefont {M.}~\bibnamefont
  {Ammosov}}, \bibinfo {author} {\bibfnamefont {N.}~\bibnamefont {Delone}}, \
  and\ \bibinfo {author} {\bibfnamefont {V.}~\bibnamefont {Krainov}},\
  }\href@noop {} {\bibfield  {journal} {\bibinfo  {journal} {Zhurnal
  Eksperimentalnoi i Teoreticheskoi Fiziki}\ }\textbf {\bibinfo {volume}
  {91}},\ \bibinfo {pages} {2008} (\bibinfo {year} {1986})}\BibitemShut
  {NoStop}%
\bibitem [{\citenamefont {Siegman}(1986)}]{siegman}%
  \BibitemOpen
  \bibfield  {author} {\bibinfo {author} {\bibfnamefont {A.~E.}\ \bibnamefont
  {Siegman}},\ }\enquote {\bibinfo {title} {Lasers},}\ \ (\bibinfo  {publisher}
  {University Science Books},\ \bibinfo {year} {1986})\ Chap.~\bibinfo
  {chapter} {20}, p.\ \bibinfo {pages} {777}\BibitemShut {NoStop}%
\bibitem [{\citenamefont {Christov}(2000)}]{2000adk}%
  \BibitemOpen
  \bibfield  {author} {\bibinfo {author} {\bibfnamefont {I.}~\bibnamefont
  {Christov}},\ }\href {\doibase 10.1364/OE.6.000034} {\bibfield  {journal}
  {\bibinfo  {journal} {Opt. Express}\ }\textbf {\bibinfo {volume} {6}},\
  \bibinfo {pages} {34} (\bibinfo {year} {2000})}\BibitemShut {NoStop}%
\bibitem [{\citenamefont {Mairesse}\ \emph {et~al.}(2003)\citenamefont
  {Mairesse}, \citenamefont {de~Bohan}, \citenamefont {Frasinski},
  \citenamefont {Merdji}, \citenamefont {Dinu}, \citenamefont {Monchicourt},
  \citenamefont {Breger}, \citenamefont {Kovacev}, \citenamefont {Taieb},
  \citenamefont {Carr\'e}, \citenamefont {Muller}, \citenamefont {Agostini},\
  and\ \citenamefont {Sali\`eres}}]{2003Mairesse}%
  \BibitemOpen
  \bibfield  {author} {\bibinfo {author} {\bibfnamefont {Y.}~\bibnamefont
  {Mairesse}}, \bibinfo {author} {\bibfnamefont {A.}~\bibnamefont {de~Bohan}},
  \bibinfo {author} {\bibfnamefont {L.~J.}\ \bibnamefont {Frasinski}}, \bibinfo
  {author} {\bibfnamefont {H.}~\bibnamefont {Merdji}}, \bibinfo {author}
  {\bibfnamefont {L.~C.}\ \bibnamefont {Dinu}}, \bibinfo {author}
  {\bibfnamefont {P.}~\bibnamefont {Monchicourt}}, \bibinfo {author}
  {\bibfnamefont {P.}~\bibnamefont {Breger}}, \bibinfo {author} {\bibfnamefont
  {M.}~\bibnamefont {Kovacev}}, \bibinfo {author} {\bibfnamefont
  {R.}~\bibnamefont {Taieb}}, \bibinfo {author} {\bibfnamefont
  {B.}~\bibnamefont {Carr\'e}}, \bibinfo {author} {\bibfnamefont {H.~G.}\
  \bibnamefont {Muller}}, \bibinfo {author} {\bibfnamefont {P.}~\bibnamefont
  {Agostini}}, \ and\ \bibinfo {author} {\bibfnamefont {P.}~\bibnamefont
  {Sali\`eres}},\ }\href {\doibase 10.1126/science.1090277} {\bibfield
  {journal} {\bibinfo  {journal} {Science}\ }\textbf {\bibinfo {volume}
  {302}},\ \bibinfo {pages} {1540} (\bibinfo {year} {2003})}\BibitemShut
  {NoStop}%
\bibitem [{\citenamefont {Gaarde}\ and\ \citenamefont
  {Schafer}(2002)}]{2002SpaceTime}%
  \BibitemOpen
  \bibfield  {author} {\bibinfo {author} {\bibfnamefont {M.~B.}\ \bibnamefont
  {Gaarde}}\ and\ \bibinfo {author} {\bibfnamefont {K.~J.}\ \bibnamefont
  {Schafer}},\ }\href {\doibase 10.1103/PhysRevLett.89.213901} {\bibfield
  {journal} {\bibinfo  {journal} {Phys. Rev. Lett.}\ }\textbf {\bibinfo
  {volume} {89}},\ \bibinfo {pages} {213901} (\bibinfo {year}
  {2002})}\BibitemShut {NoStop}%
\bibitem [{\citenamefont {Miao}\ \emph {et~al.}(2012)\citenamefont {Miao},
  \citenamefont {Zeng}, \citenamefont {Liu}, \citenamefont {Zheng},
  \citenamefont {Li}, \citenamefont {Xu}, \citenamefont {Platonenko},\ and\
  \citenamefont {Strelkov}}]{Miao2012}%
  \BibitemOpen
  \bibfield  {author} {\bibinfo {author} {\bibfnamefont {J.}~\bibnamefont
  {Miao}}, \bibinfo {author} {\bibfnamefont {Z.}~\bibnamefont {Zeng}}, \bibinfo
  {author} {\bibfnamefont {P.}~\bibnamefont {Liu}}, \bibinfo {author}
  {\bibfnamefont {Y.}~\bibnamefont {Zheng}}, \bibinfo {author} {\bibfnamefont
  {R.}~\bibnamefont {Li}}, \bibinfo {author} {\bibfnamefont {Z.}~\bibnamefont
  {Xu}}, \bibinfo {author} {\bibfnamefont {V.~T.}\ \bibnamefont {Platonenko}},
  \ and\ \bibinfo {author} {\bibfnamefont {V.~V.}\ \bibnamefont {Strelkov}},\
  }\href {\doibase 10.1364/OE.20.005196} {\bibfield  {journal} {\bibinfo
  {journal} {Opt. Express}\ }\textbf {\bibinfo {volume} {20}},\ \bibinfo
  {pages} {5196} (\bibinfo {year} {2012})}\BibitemShut {NoStop}%
\bibitem [{\citenamefont {Chang}(2005)}]{chang}%
  \BibitemOpen
  \bibfield  {author} {\bibinfo {author} {\bibfnamefont {Z.}~\bibnamefont
  {Chang}},\ }\href {\doibase 10.1103/PhysRevA.71.023813} {\bibfield  {journal}
  {\bibinfo  {journal} {Phys. Rev. A}\ }\textbf {\bibinfo {volume} {71}},\
  \bibinfo {pages} {023813} (\bibinfo {year} {2005})}\BibitemShut {NoStop}%
\end{thebibliography}%

\end{document}